\documentclass[useAMS,usenatbib]{mn2e}

\usepackage{graphicx}
\usepackage{times}
\usepackage{natbib}

\renewcommand{\vec}[1]{ {\bmath #1} }

\title[Turbulent gas motions in galaxy cluster simulations: The role of SPH
viscosity] {Turbulent gas motions in galaxy cluster simulations: The role of
  SPH viscosity} \author[K. Dolag, F. Vazza, G. Brunetti, G. Tormen]
{K.Dolag$^{1}$,$^{2}$\thanks{E-mail: kdolag@mpa-garching.mpg.de}, F.
  Vazza$^{1}$, G. Brunetti$^{3}$, G. Tormen$^{1}$\\
  $^{1}$Dipartimento di Astronomia, Universita di Padova, vicolo
  dell'Osservatorio 2, 35122 Padova, Italy \\
  $^{2}$ Max-Planck-Institut f\"ur Astrophysik, Garching, Germany\\
  $^{3}$ INAF, Istituto di Radioastronomia, via P.Gobetti 101, 40129 Bologna,
  Italy}

\begin{document}

\date{Accepted ???. Received ???; in original form ???}

\pagerange{\pageref{firstpage}--\pageref{lastpage}} \pubyear{0000}

\maketitle

\label{firstpage}

\begin{abstract}
  Smoothed particle hydrodynamics (SPH) employs an artificial viscosity to
  properly capture hydrodynamical shock waves. In its original formulation,
  the resulting numerical viscosity is large enough to suppress structure in
  the velocity field on scales well above the nominal resolution limit, and to
  damp the generation of turbulence by fluid instabilities. This could
  artificially suppress random gas motions in the intracluster medium (ICM),
  which are driven by infalling structures during the hierarchical structure
  formation process. We show that this is indeed the case by analysing results
  obtained with an SPH formulation where an individual, time-variable
  viscosity is used for each particle, following a suggestion by
  \citet{1997JCoPh..136....41S}.  Using test calculations involving strong
  shocks, we demonstrate that this scheme captures shocks as well as the
  original formulation of SPH, but, in regions away from shocks, the numerical
  viscosity is much smaller.  In a set of nine high-resolution simulations of
  cosmological galaxy cluster formation, we find that this low--viscosity
  formulation of SPH produces substantially higher levels of turbulent gas
  motions in the ICM, reaching a kinetic energy content in random gas motions
  (measured within a 1Mpc cube) of up to 5\%-30\% of the thermal energy
  content, depending on cluster mass.  This has also significant effects on
  radial gas profiles and bulk cluster properties.  We find a central
  flattening of the entropy profile and a reduction of the central gas density
  in the low--viscosity scheme. As a consequence, the bolometric X-ray
  luminosity is decreased by about a factor of two. However, the cluster
  temperature profile remains essentially unchanged. Interestingly, this tends
  to reduce the differences seen in SPH and adaptive mesh refinement
  simulations of cluster formation. Finally, invoking a model for particle
  acceleration by MHD waves driven by turbulence, we find that efficient
  electron acceleration and thus diffuse radio emission can be powered in the
  clusters simulated with the low viscosity scheme provided that more than
  5\%-10\% of the turbulent energy density is associated with Fast
  Magnetosonic Modes.
\end{abstract}



\section{Introduction} \label{sec:intro}

In hierarchical cold dark matter cosmologies, large structures form through
the accretion of smaller structures \citep[e.g.,][]{1993MNRAS.261L...8W}. In
particular, mergers and infall of halos play a fundamental role in determining
the structure and dynamics of massive clusters of galaxies, where mergers can
induce large--scale bulk motions with velocities of the order of $\sim 1000$ km
s$^{-1}$ or larger. This results in complex hydrodynamic flows where most of
the kinetic energy is quickly dissipated to heat by shocks, but some part may
in principle also excite long-lasting turbulent gas motions.

Numerical simulations of merging clusters
\citep[e.g.,][]{1993A&A...272..137S,1997ApJS..109..307R,2001ApJ...561..621R,
  2005astro.ph..5274T} provide a detailed description of the gas-dynamics
during a merger event.  It has been shown that infalling sub-clusters can
generate laminar bulk flows through the primary cluster and inject turbulent
eddies via Kelvin-Helmholtz (KH) instabilities at interfaces between the bulk
flows and the primary cluster gas. Such eddies redistribute the energy of the
merger through the cluster volume in a few turnover times, which corresponds
to a time interval of order 1 Gyr. The largest eddies decay with time into
more random and turbulent velocity fields, eventually developing a turbulent
cascade with a spectrum of fluctuations expected to be close to a Kolmogorov
spectrum.

Observationally, spatially resolved gas pressure maps of the Coma cluster
obtained from a mosaic of XMM--Newton observation have indeed revealed the
signature of mildly supersonic turbulence, at least in the central regions of
the cluster \citep{2004A&A...426..387S}.  It has also been suggested that the
micro--calorimeters on-board of future X--ray satellites such as ASTRO-E2
should be able to detect the turbulent broadening of the lines of heavy ions
in excess of the thermal broadening \citep{2003AstL...29..791I}, which would
represent a direct measurement of cluster turbulence.

Cluster turbulence could in principle store an appreciable fraction of the
thermal energy of massive clusters, which would make it an important factor
for understanding the structure of the ICM.  Shear flows associated with
cluster turbulence and the resulting dynamo processes could considerably
amplify the magnetic field strength in the ICM
\citep[e.g.,][]{1999A&A...348..351D,Dolag:2002}.  In addition,
magnetohydrodynamic waves can be efficiently injected in the ICM directly by
shocks, by Kelvin-Helmholtz or Rayleigh-Taylor instabilities, or by the decay
of turbulent eddies at larger scales.  These waves, as well as shocks, can
efficiently accelerate supra--thermal particles in the ICM to higher energies.
Although there is still some debate concerning the detailed mechanism
responsible for the origin of relativistic particles and magnetic fields in
the ICM \citep[e.g.,][]{2003mecg.conf..349B}, the presence of relativistic
electrons and of $\sim \mu$G--strength magnetic fields in the ICM is proven by
non--thermal emission studied with radio observations and possibly
observations of hard X-Ray emission \citep[e.g.,][for a
review]{2002mpgc.book.....F,2003A&A...398..441F}.  In addition, the occurrence
of non--thermal phenomena is found to be related to the dynamical state and
mass of the parent cluster
\citep{1999NewA....4..141G,buote2001,2001A&A...378..408S,2002IAUS..199..133F},
which suggests a connection between cluster mergers and non--thermal activity.

Despite this potentially significant relevance of turbulence for the ICM, 
  quantitative studies have received comparatively little attention in
hydrodynamical cluster simulations thus far. One reason for this is that 3D
turbulence is difficult to resolve in any numerical scheme, because
these always introduce some finite numerical viscosity, 
effectively putting a limit
on the Reynolds numbers that can still be adequately represented.  In the
Lagrangian SPH method, which has been widely employed for studies of cluster
formation, an artificial viscosity is used to capture shocks. The 
original
parameterisation of this viscosity \citep{1983JCoPh..52....374S} makes the
scheme comparatively viscous; it smoothes out small-scale velocity
fluctuations and viscously damps random gas motions well above the nominal
resolution limit. This hampers the ability of the original SPH 
to develop fluid turbulence down to the smallest resolved scales.

However, the numerical viscosity of SPH can in principle be reduced by using a
more sophisticated parameterisation of the artificial viscosity. Ideally, the
viscosity should only be present in a hydrodynamical shock, but otherwise it
should be negligibly small.  To come closer to this goal,
\citet{1997JCoPh..136....41S} proposed a numerical scheme where the artificial
viscosity is treated as an independent dynamical variable for each particle,
with a source term triggered by shocks, and an evolutionary term that lets the
viscosity decay in regions  away from shocks.  In this way, one can hope that shocks can
still be captured properly, while in the bulk of the volume of a simulation,
the effective viscosity is lower than in original SPH. We adopt this scheme
and implement it in a cosmological simulation code. We then apply it to
high-resolution simulations of galaxy cluster formation, allowing us to
examine a better representation of turbulent gas motions in SPH simulations of
clusters.  This also shines new light on differences in the results of
cosmological simulations between different numerical techniques.

In Section~\ref{sec:code}, we discuss different ways of implementing the
artificial viscosity in SPH. We demonstrate in Section~\ref{sec:tests} the
robustness of our new low--viscosity scheme by applying it to several test
problems. In Sections~\ref{sec:simulations}, \ref{sec:turbulence} and
\ref{sec:cluster}, we introduce our set of cluster simulations, the algorithm
to detect and measure turbulence, and the implications of the presence of
turbulence for the structure and properties of galaxy clusters.  In
Section~\ref{sec:lines}, we consider the effects of turbulence on the
line-width of narrow X-ray metal lines.  Finally, in Section~\ref{sec:radio}
we apply the results from our new simulations to models for the production of
radio emission due to turbulent acceleration processes. We give our
conclusions in Section~\ref{sec:conclusions}.


\section{Simulation Method} \label{sec:code}

The smoothed particle hydrodynamics method treats shocks with an artificial
viscosity, which leads to a broadening of shocks and a relatively rapid
vorticity decay.  To overcome these problems, \citet{1997JCoPh..136....41S}
proposed a new parameterisation of the artificial viscosity capable of
reducing the viscosity in regions away from shocks, where it is not needed,
while still being able to capture strong shocks reliably.
We have implemented this method in the cosmological SPH code {\small GADGET-2}
\citep{2005astro.ph..5010S}, and describe the relevant details in the
following. 

In {\small GADGET-2}, the viscous force is implemented as
\begin{equation}
  \frac{{\mathrm d}v_i}{{\mathrm d}t} = - 
\sum_{j=1}^{N}m_j\Pi_{ij}\nabla_i\bar{W}_{ij},
\end{equation}
and the rate of entropy change due to viscosity is
\begin{equation}
  \frac{{\mathrm d}A_i}{{\mathrm d}t} = - 
\frac{1}{2}\frac{\gamma-1}{\rho_i^{\gamma-1}}
\sum_{j=1}^{N}m_j\Pi_{ij}v_{ij}\cdot\nabla_i\bar{W}_{ij},
\end{equation}
where $A_i=(\gamma-1)u_i/\rho_i^{\gamma-1}$ is the entropic function of a
particle of density $\rho_i$ and thermal energy $u_i$ per unit mass, and
$\bar{W}_{ij}$ denotes the arithmetic mean of the two kernels $W_{ij}(h_i)$
and $W_{ij}(h_j)$.
The usual
parameterisation of the artificial viscosity
\citep{1983JCoPh..52....374S,1995JCoPh.121..357B} for an interaction of two
particles $i$ and $j$ includes terms to mimic a shear and bulk viscosity.
For standard cosmological
SPH simulations, it can be written as
\begin{equation}
\Pi_{ij}=\frac{-\alpha c_{ij}\mu_{ij}+\beta\mu_{ij}^2}{\rho_{ij}}f_{ij},
\label{eqn:visc}
\end{equation}
for $\vec{r}_{ij}\cdot\vec{v}_{ij}\le 0$ and $\Pi_{ij} = 0$ otherwise,
i.e.~the pair-wise viscosity is only non-zero if the particle are approaching
each other.  Here
\begin{equation}
\mu_{ij}=\frac{h_{ij}\vec{v}_{ij}\cdot\vec{r}_{ij}
}{\vec{r}_{ij}^2+\eta^2},
\end{equation}
$c_{ij}$ is the arithmetic mean of the two sound speeds, $\rho_{ij}$ is the
average of the densities, $h_{ij}$ is the arithmetic mean of the smoothing
lengths, and $\vec{r}_{ij}=\vec{r}_i-\vec{r}_j$ and
$\vec{v}_{ij}=\vec{v}_i-\vec{v}_j$ are the inter-particle distance and
relative velocity, respectively. We have also included a viscosity-limiter
$f_{ij}$, which is often used to suppress the viscosity locally in regions of
strong shear flows, as measured by
\begin{equation}
f_i=\frac{|\left<\vec{\nabla}\cdot\vec{v}\right>_i|}{|\left<\vec{\nabla}\cdot\vec{v}\right>_i| +
|\left<\vec{\nabla}\times\vec{v}\right >_i|+\sigma_i},
\end{equation}
which can help to avoid spurious angular momentum and vorticity transport in
gas disks \citep{1996IAUS..171..259S}. Note however that the parameters
describing the viscosity (with common choices $\alpha=0.75-1.0$,
$\beta=2\alpha$, $\eta=0.01 h_{ij}$, and $\sigma_i=0.0001 c_i/h_i$ ) stay here
fixed in time.  This then defines the `original' viscosity scheme usually
employed in cosmological SPH simulations. We refer to runs performed with this
viscosity scheme as {\it ovisc} simulations.

As a variant of the original parameterisation of the 
artificial viscosity,
{\small GADGET-2} can use a formulation proposed by
\cite{1997JCoPh..136....298S} based on an analogy with Riemann solutions of
compressible gas dynamics.  In this case, $\mu_{ab}$ is defined as
\begin{equation}
\mu_{ij}=\frac{\vec{v}_{ij}\cdot\vec{r}_{ij}
}{|\vec{r}_{ij}|},
\end{equation}
and one introduces a
 signal velocity $v_{ij}^{\rm sig}$, for example in the form
\begin{equation}
v_{ij}^{\rm sig} = c_i + c_j - 3\mu_{ij}.
\end{equation}
The resulting viscosity term then changes into
\begin{equation}
\Pi_{ij}=\frac{-0.5\alpha v_{ij}^{sig} \mu_{ij}}{\rho_{ij}}f_{ij}.
\label{eqn:visc2}
\end{equation}
We have also performed simulations using this signal velocity based artificial
viscosity and found that it performs well in all test problems we examined so
far, while in some cases it performed slightly better, in particular avoiding
post shock oscillations in a more robust way. We refer to simulations
performed using this `signal velocity' based viscosity scheme as {\it svisc}
simulations.

The idea proposed by \cite{1997JCoPh..136....41S} is to give every particle
its own viscosity parameter $\alpha_i$, which is allowed to evolve with time
according to
\begin{equation}
\frac{{\rm d}\alpha_i}{{\rm d}t}=-\frac{\alpha_i-\alpha_{\rm min}}{\tau}+S_i.
\end{equation}
This causes $\alpha_i$ to decay to a minimum value $\alpha_{\rm min}$ with an
e-folding time $\tau$, while the source term $S_i$ is meant to make $\alpha_i$
rapidly grow when a particle approaches a shock. For the decay timescale,
\cite{1997JCoPh..136....41S} proposed to use
\begin{equation}
\tau=h_{i}\,/\,(c_i\,l),
\end{equation}
where $h_i$ is the smoothing length, $c_{i}$ the sound speed and $l$ a free
 (dimensionless) parameter which determines on how many information crossing times the
viscosity decays. For an ideal gas and a strong
shock, this time scale can be related to a length scale $\delta=0.447/l$ (in
units of the smoothing length $h_i$) on which the viscosity parameter decays
behind the shock front. For the source term $S_i$, we follow
\cite{1997JCoPh..136....41S} and adopt
\begin{equation}
S_i = S^* f_i\,\, \mathrm{max}(0,-|\left<\vec{\nabla}\cdot\vec{v}\right>_i|),
\end{equation}
where $\left<\vec{\nabla}\cdot\vec{v}\right>_i$ denotes the usual SPH estimate
of the divergence around the particle $i$. Note that it would in principle be
possible to use more sophisticated shock detection schemes here, but the
simple criterion based on the convergence of the flow is already working well
in most cases. We refer to simulations carried our with this `low'
viscosity scheme as {\it lvisc} runs.

Usually we set $S^* = 0.75$ and choose $l=1$.  We also
restrict $\alpha_i$ to be in the range $\alpha_{\rm min}=0.01$ and
$\alpha_{\rm max}=0.75$. Choosing $\alpha_{\rm min}>0$ has the
advantage, that possible noise which might be present in the 
velocity representation by the particles on scales below the smoothing 
length still will get damped with time.
Increasing $S^*$ can give a faster response of the
artificial viscosity to the shock switch without inducing higher viscosity
than necessary elsewhere. We also note that we replace $\alpha$ in equation
\ref{eqn:visc} (and equation \ref{eqn:visc2} respectively)
by the arithmetic mean $\alpha_{ij}$ of two interacting
particles.  Depending on the problem, we initialise $\alpha_i$ at the start of
a simulation either with $\alpha_{\rm min}$ or $\alpha_{\rm max}$, depending
on whether or not there are already shocks present in the initial conditions,
respectively.

While the approach to reduce numerical
viscosity with a time-variable $\alpha_i$ works well with both basic
parameterisations of the artificial viscosity, most of our cosmological
simulations were carried out with the 'original' parameterisation because the
signal velocity variant became available in {\small GADGET-2} only recently.


\begin{figure}
\includegraphics[width=0.49\textwidth]{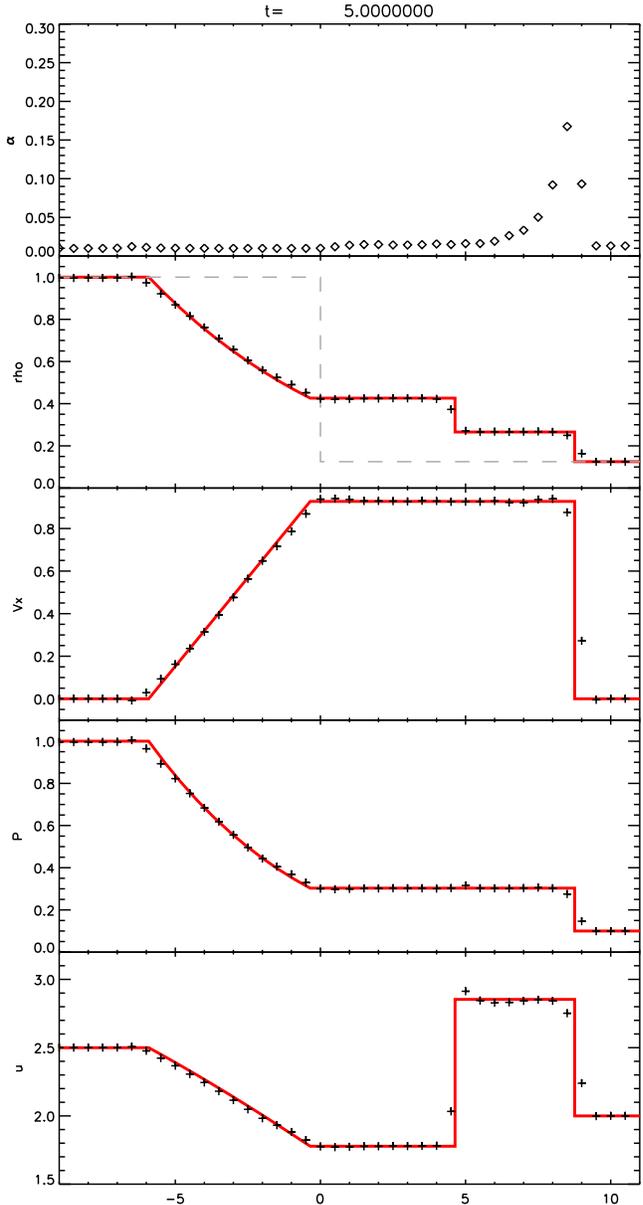}
\caption{A standard shock tube problem \citep{1978JCoPh..27....1S}
computed with the low--viscosity scheme with an individual,
time-dependent viscosity. From top to bottom, we show  current
value of the strength of the artificial viscosity $\alpha_i$,
density, velocity, pressure, and internal energy, averaged for
bins with spacing equal to the SPH smoothing length for particles
in the low density region. The analytic solution of the problem
for the time $t=5.0$ is shown as a solid line.} \label{fig:sod}
\end{figure}

\begin{figure*}
\includegraphics[width=1.0\textwidth]{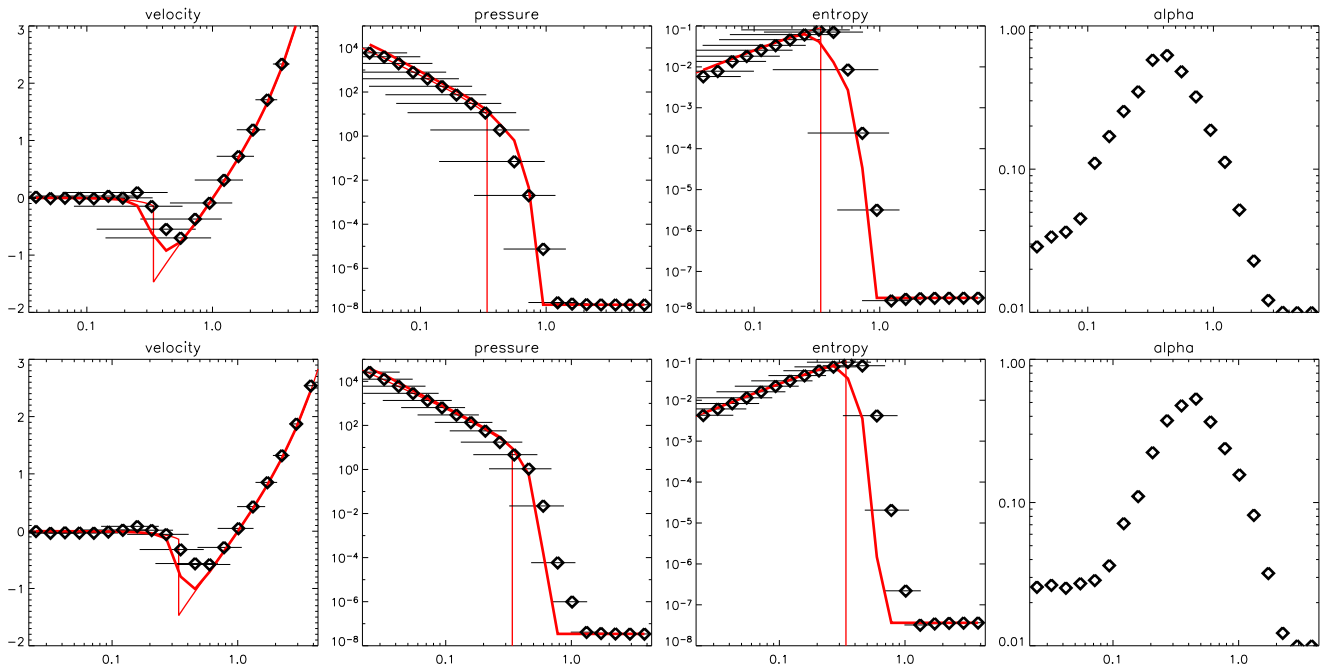}
\caption{Profiles of velocity (left column), pressure (middle left column)
  entropy (middle right column) and viscosity constant $\alpha$ (right column) for
  the spherical collapse test at 2 different times (from top to bottom). The
  thin line marks the analytic solution, diamonds give the
  result obtained by the new SPH formulation for the time dependent
  viscosity. The thick line is the analytic solution adaptively
  smoothed with the SPH kernel, using the smoothing length of the
  particles in each bin. The lengths of the horizontal lines plotted at each
  data point correspond to the smoothing lengths of the SPH particles
  at this position.
} \label{fig:bert1}
\end{figure*}

\section{Test Simulations} \label{sec:tests}

To verify that the low--viscosity formulation of SPH with its time-dependent
artificial viscosity is still able to correctly capture strong shocks, we
computed a number of test problems. We here report on a standard shock tube
test, and a spherical collapse test, which both have direct relevance for the
cosmological formation of halos. As a more advanced test for the ability of
the code to follow vorticity generation, we investigate the problem of a
strong shock striking an overdense cloud in a background medium. This test can
also give hints whether turbulence is less strongly suppressed in the
low--viscosity treatment of SPH than in the original formulation.

\subsection{Shock-Tube Test}

First, we computed a shock tube problem, which is a common test for
hydrodynamical numerical schemes \citep{1978JCoPh..27....1S}. For
definiteness, a tube is divided into two halves, having density $\rho_1=1$ and
pressure $p_1=1$ on the left side, and $\rho_2=0.125$ and $p_2=0.1$ on the
right side, respectively.  Like in Sod (1978), we assume an ideal gas with
$\gamma=1.4$. To avoid oscillations in the post shock region (note that a
shock is present in the initial conditions) we initialise the viscosity values
of the particles with $\alpha_{\rm max}=0.75$. We compute the test in 3D and
make use of periodic boundary conditions. The initial particle grid is 5x5x50
on one half, and 10x10x100 on the other half, implying an equal particle mass
for both sides.

In Figure \ref{fig:sod}, we show the state of the system at simulation
time $t=5$ in terms of density, velocity, and internal energy with a
binning which corresponds to the smoothing length for particles in the
low density region.  We also include the analytic expectation for
comparison. In addition, we plot the values of the artificial
viscosity parameter of the particles. Clearly visible is that the
viscosity is close to $\alpha_{\rm min}$ everywhere, except in the
region close to the shock. One can also see how the viscosity builds
up to its maximum value in the pre-shock region and decays in the
post shock region.
We note that the final post-shock state of the gas
agrees well with the theoretical expectation, and is indistinguishable
from the case where the original viscosity parameterisation is used.

\begin{figure*}
\includegraphics[width=0.49\textwidth]{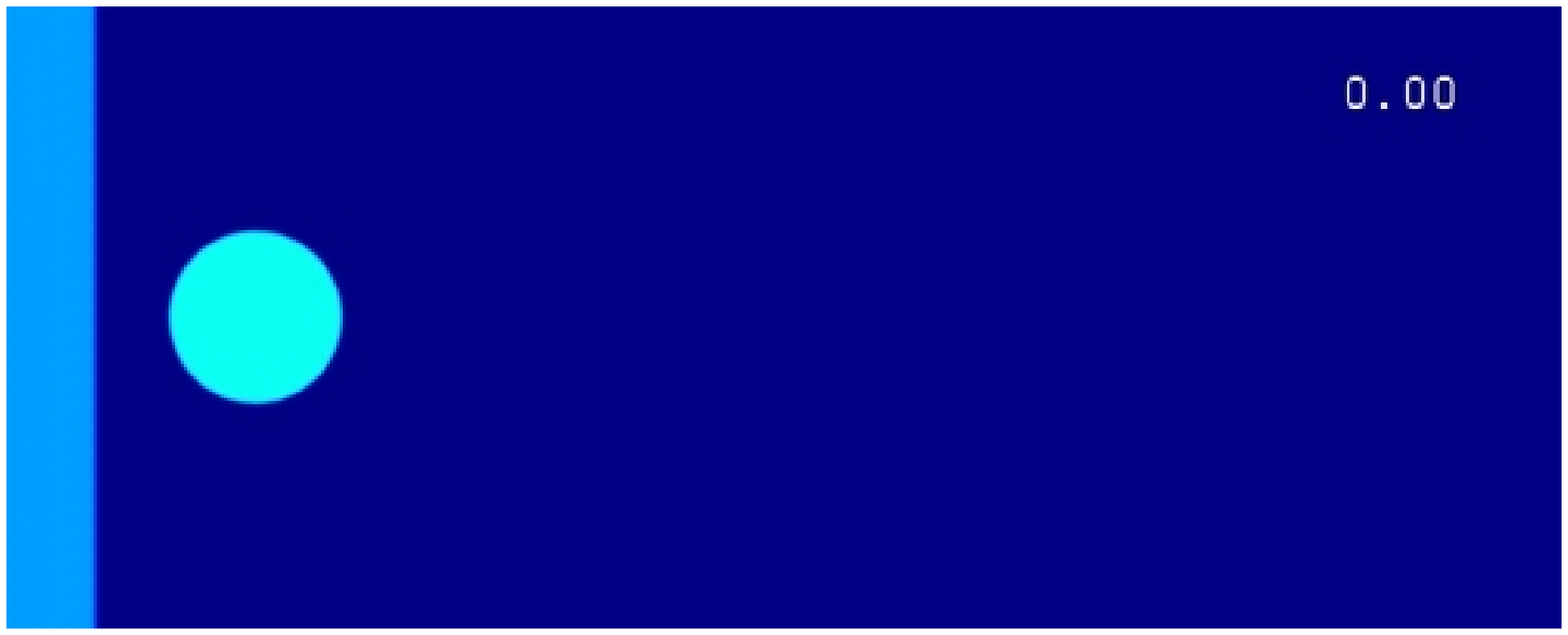}
\includegraphics[width=0.49\textwidth]{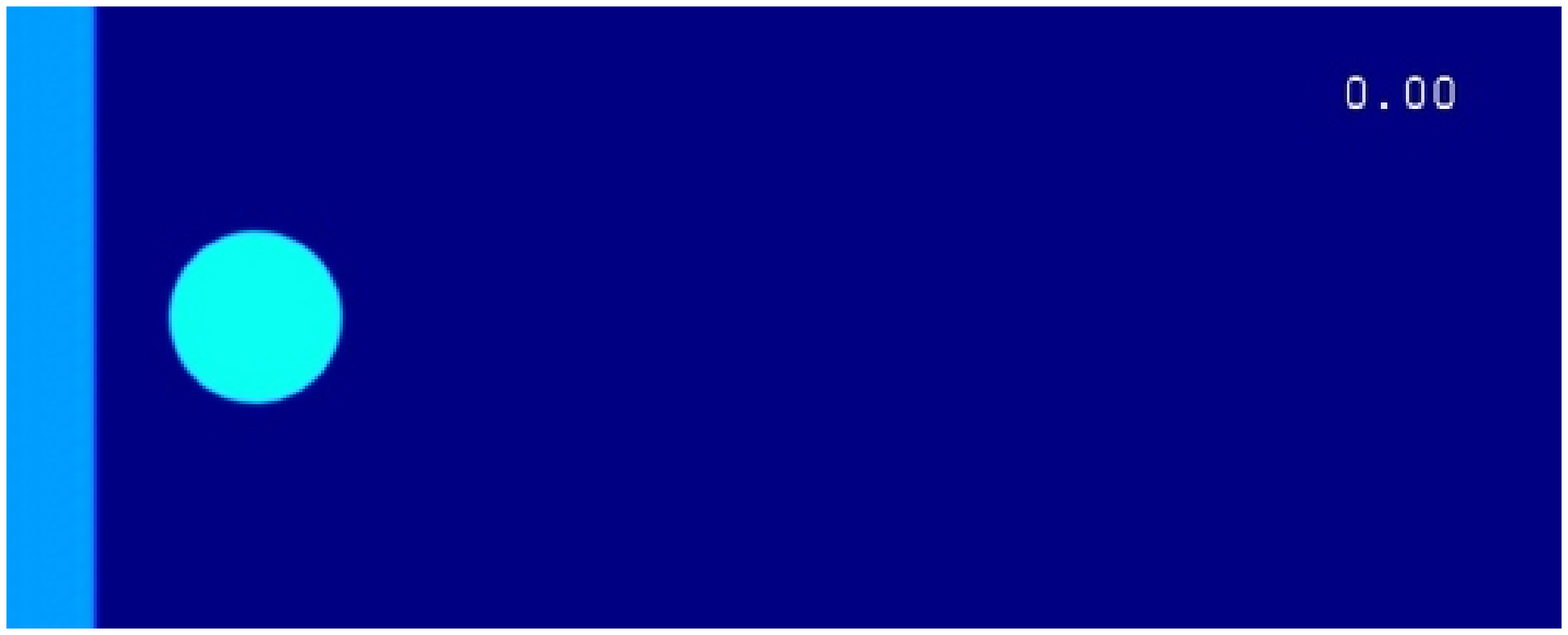}\\
\includegraphics[width=0.49\textwidth]{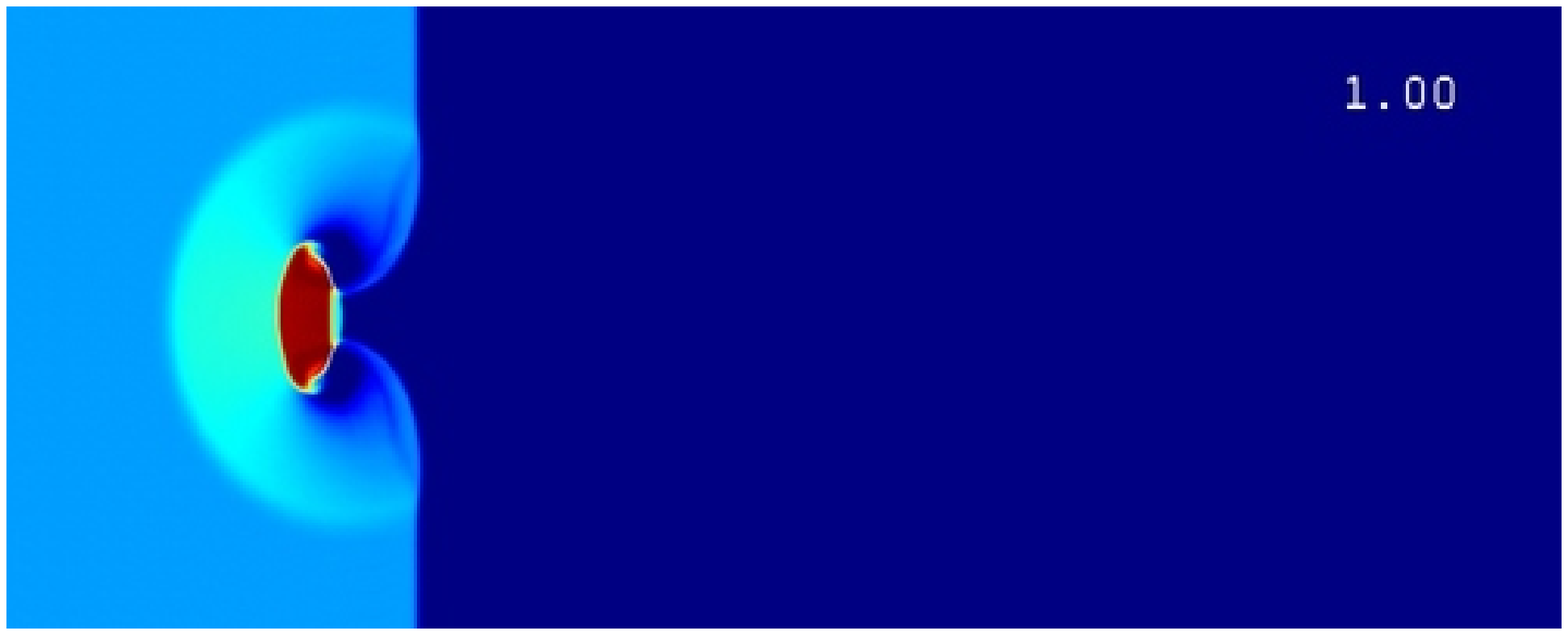}
\includegraphics[width=0.49\textwidth]{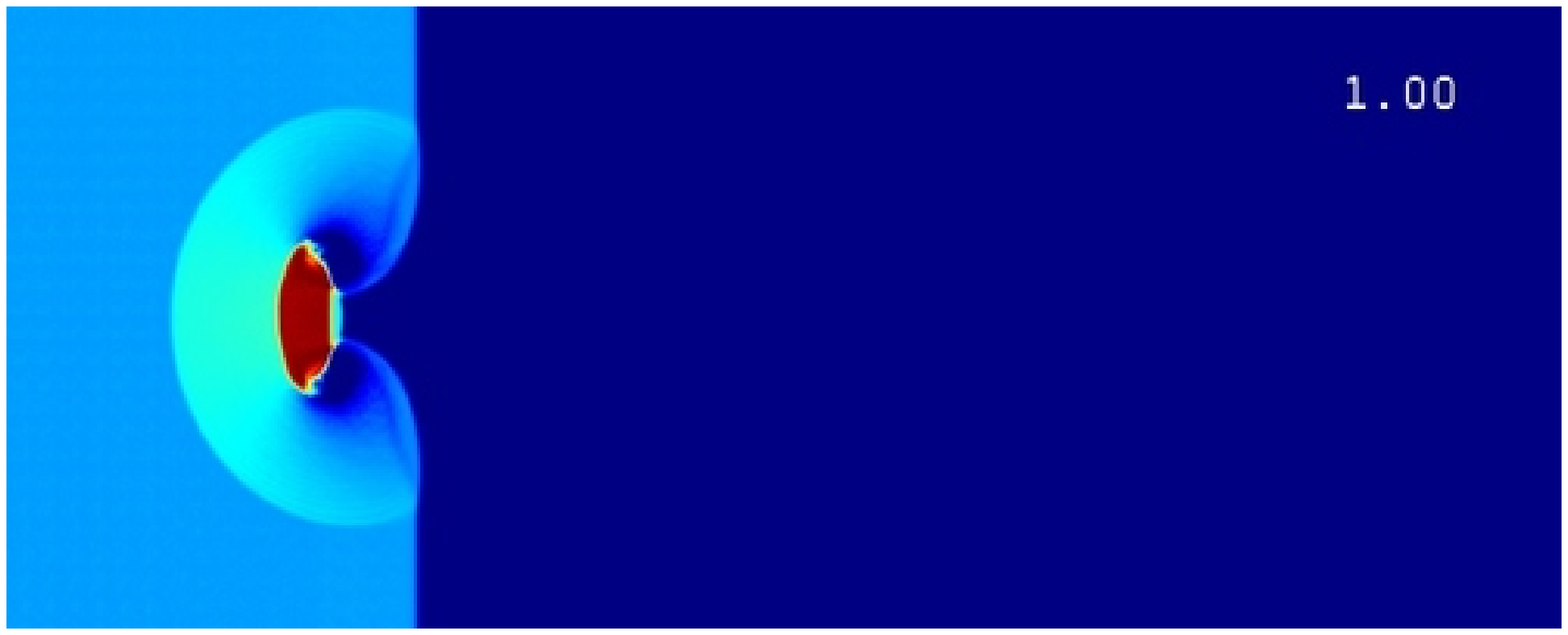}\\
\includegraphics[width=0.49\textwidth]{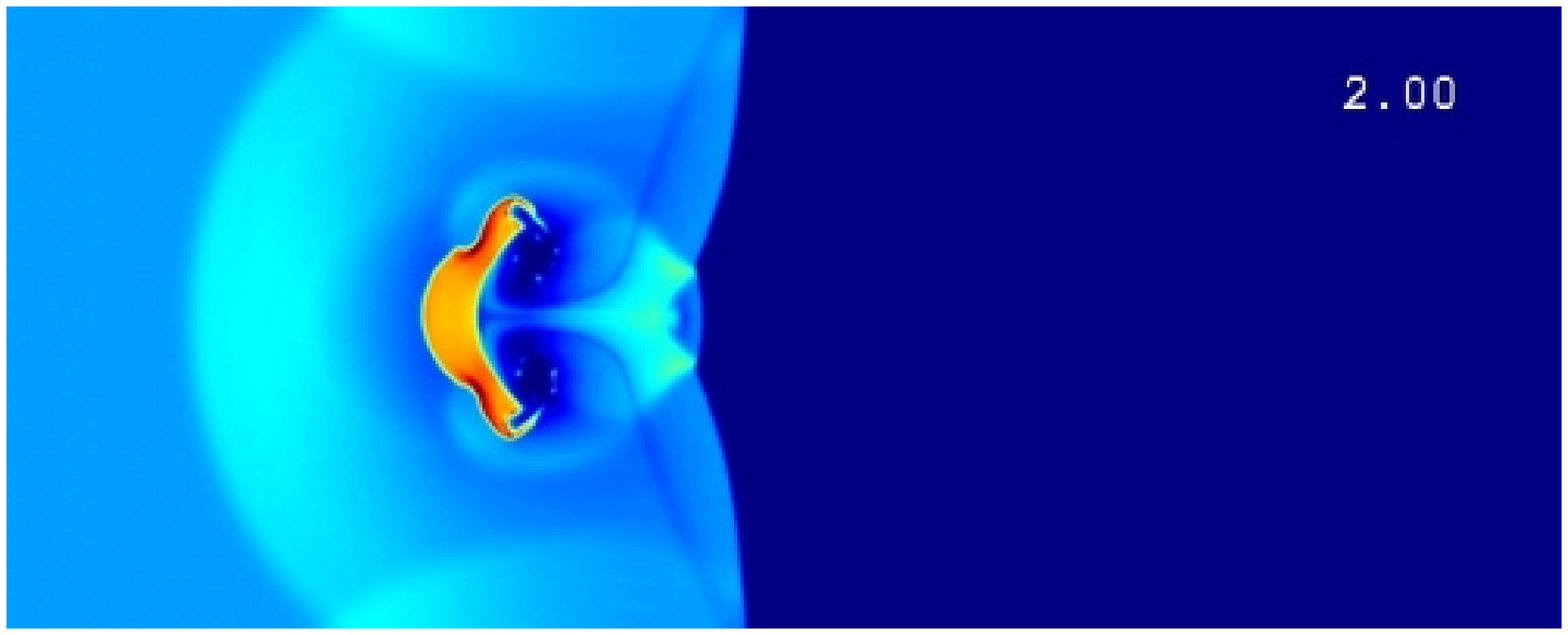}
\includegraphics[width=0.49\textwidth]{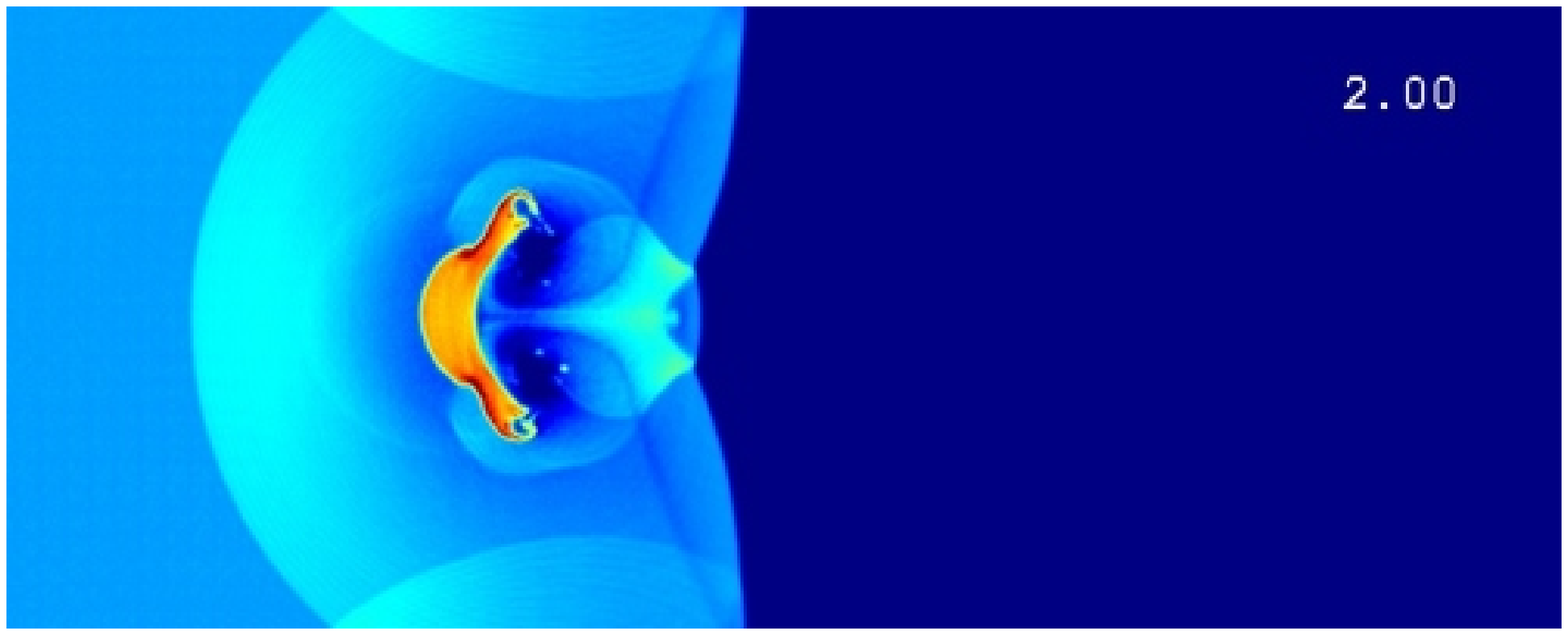}\\
\includegraphics[width=0.49\textwidth]{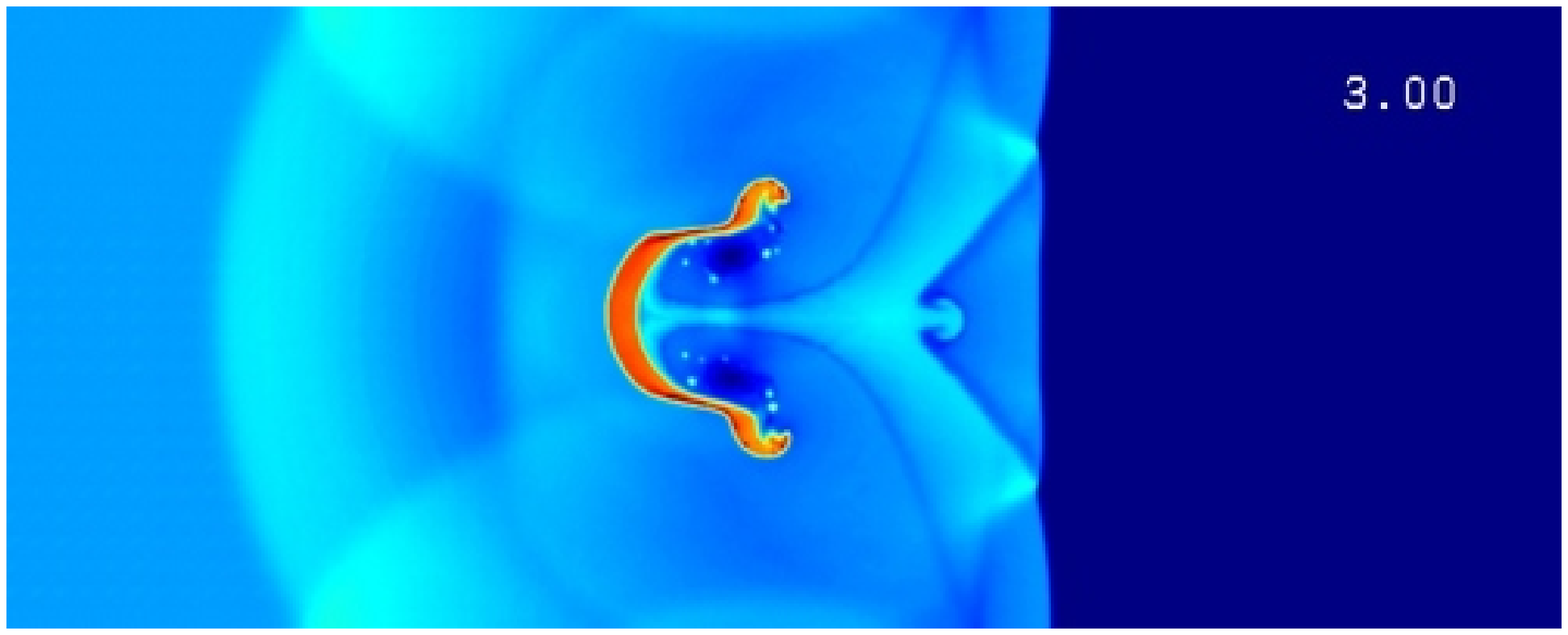}
\includegraphics[width=0.49\textwidth]{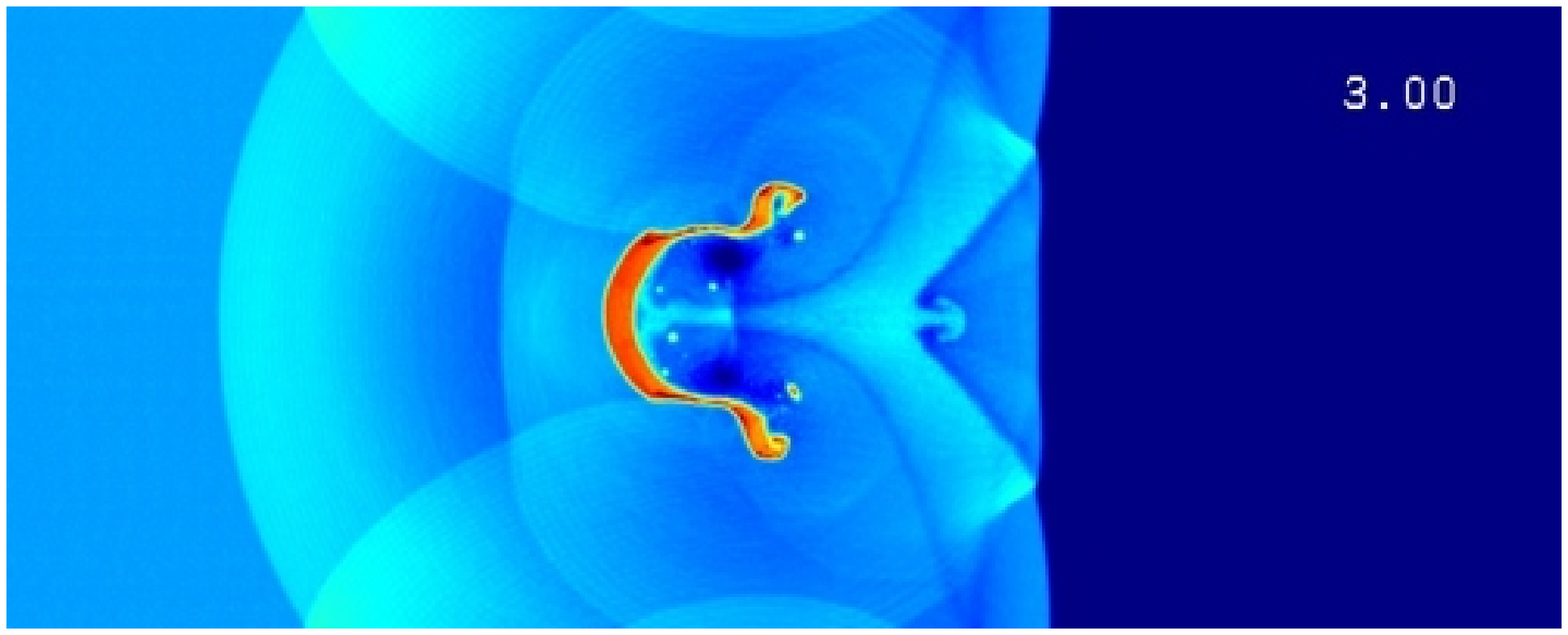}\\
\includegraphics[width=0.49\textwidth]{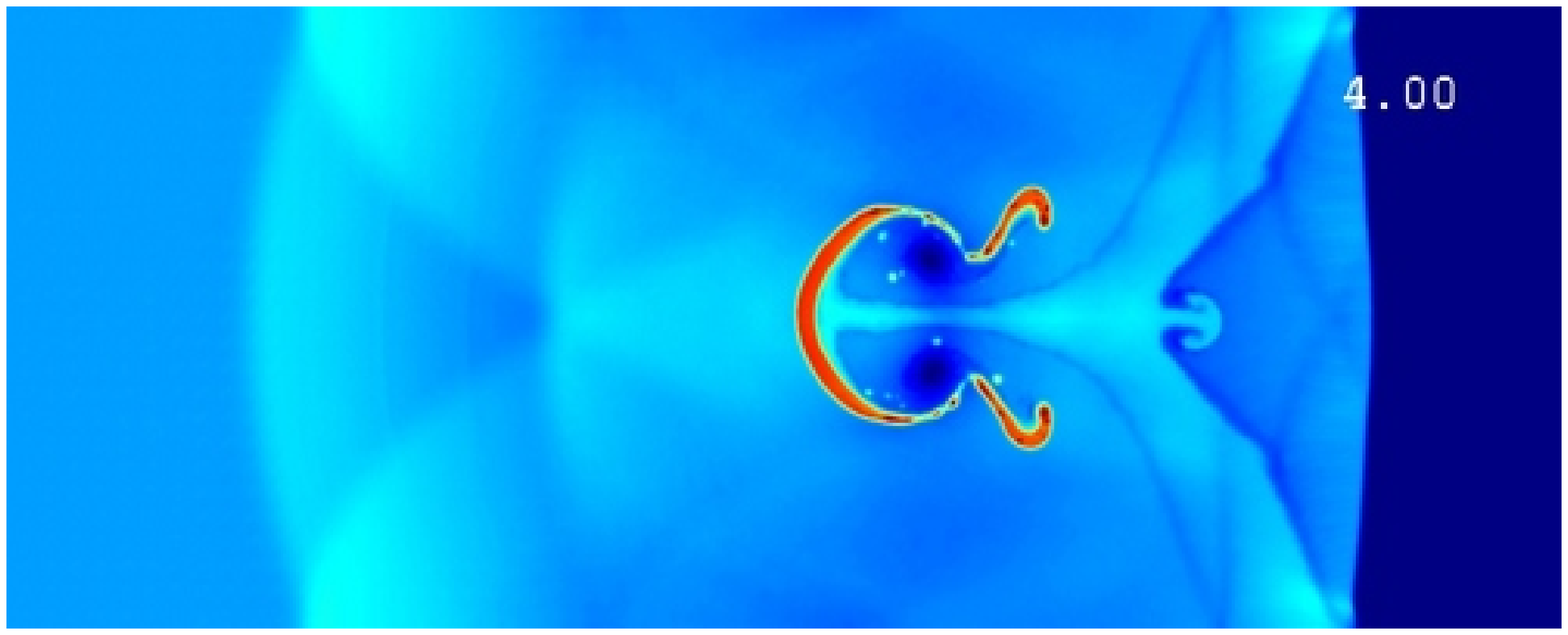}
\includegraphics[width=0.49\textwidth]{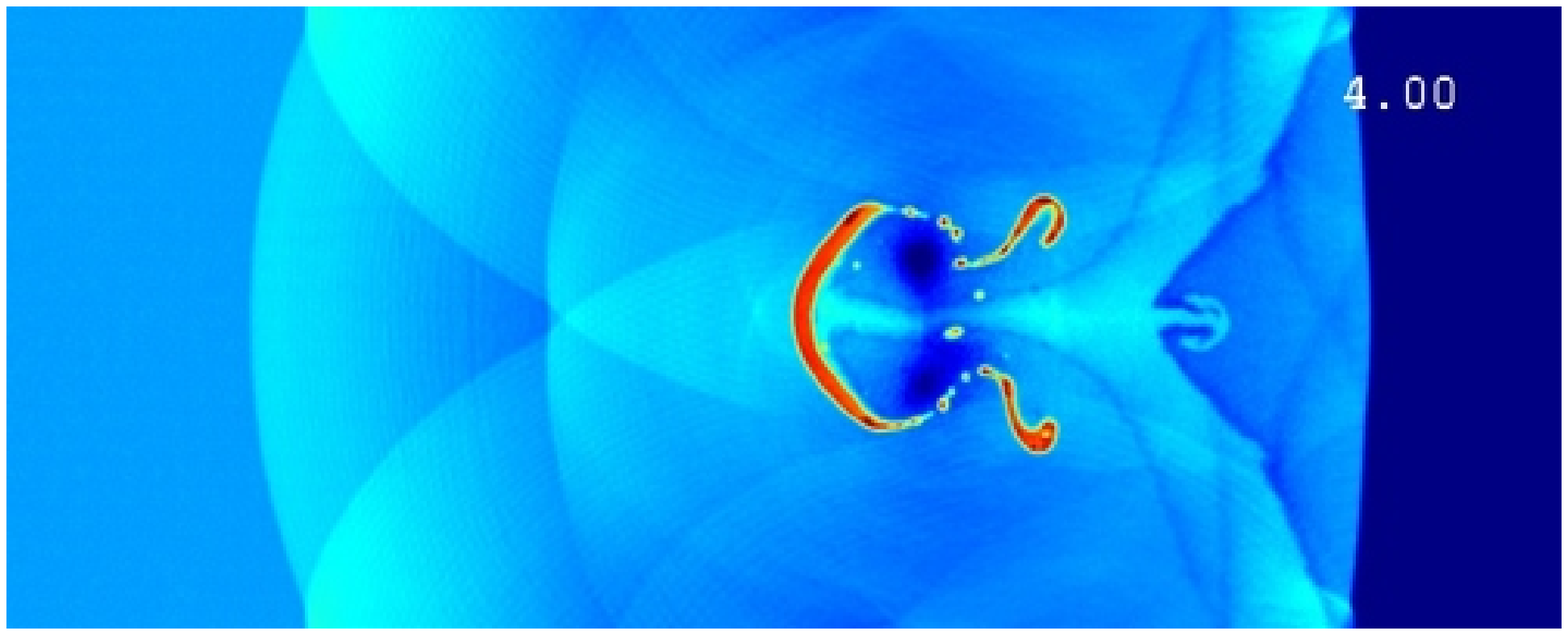}\\
\caption{Time evolution of the interaction of a strong shock wave with an
  overdense cloud. We show the projected gas density and compare simulations
  carried out with original SPH (left) and the low--viscosity formulation
  (right).  The incident shock wave has Mach number 10, and the cloud is
  initially at pressure equilibrium in the ambient medium and has overdensity
  5.}
\label{fig:tube2d}
\end{figure*}

\subsection{Self-similar spherical collapse}

A test arguably more relevant for cosmological structure formation is the
self-similar, spherical collapse of a point perturbation in a homogeneous
expanding background \citep{1985ApJS...58....1B}. This test is difficult for
grid and SPH codes alike. The gas cloud collapses self similarly, forming a
very strong shock (with formally has infinite Mach number) at the accretion
surface. While grid codes with shock capturing schemes can usually recover the
sharp shock surface very well, the large dynamic range in the post shock
region with its singular density cusp, as well as the strict spherical
symmetry of the problem, are challenging for mesh codes.  On the
other hand, Lagrangian SPH codes tend to have fewer problems with the central
structure of the post-shock cloud, but they broaden the shock surface
substantially, and typically show appreciable pre-shock entropy injection as
result of the artificial viscosity.

We have computed the self-similar collapse test and compared the results for
the new viscosity parameterisation with the analytic expectation.  The very
strong spherical shock of this problem is a particularly interesting test,
because we can here test whether the low--viscosity formulation is still able
to capture the strongest shocks possible.

In Figure \ref{fig:bert1}, we show the structure of the shock at 2 consecutive
times, scaled to the self-similar variables. In general, the SPH result
recovers the analytic solution for the post-shock state very well, especially
when the entropy profile is considered. However, the shock is substantially
broadened, and some pre-heating in front of the shock is clearly visible. In
the velocity field, some weak post-shock oscillations are noticable. We have
also indicated the smoothing lengths of the SPH particles as horizontal error
bars for each of the data points (the points at which the SPH kernel falls to
zero is reached at twice this length). For comparison, we additionally
over-plotted the analytic solution adaptively smoothed with the SPH kernel
size at each bin.

The panels of the right column in Figure \ref{fig:bert1} show the profile of
the viscosity parameter, which was set to $\alpha_{\rm min}$ at the beginning
of the simulation, as the initial conditions do not contain a shock. The
viscosity parameter builds up immediately after starting the simulation as the
strong shock forms. Later one can see how the viscosity parameter begins to
decay towards $\alpha_{\rm min}$ in the inner part, how it builds up to
$\alpha_{\rm max}$ towards the shock surface, and how a characteristic profile
develops as the shock moves outward. In the post-shock region an intermediate
viscosity values is maintained for some time due to some non-radial motions of
gas particles in this region.

\begin{table*}non radiative 
\caption{Main characteristics of the non radiative galaxy cluster simulations.
Column 1: identification label. Columns 2 and 3:
mass of the dark matter ($M_{\rm DM}$) and gas ($M_{\rm gas}$)
components inside the virial radius.  Column 4: virial radius
$R_{\rm v}$. Column 5: X-ray luminosity inside the virial radius $L_x$.
Columns 6 and 7: mass-weighted temperature ($T_{\rm MW}$) and
spectroscopic like temperature ($T_{\rm SL}$).}
\begin{center}
\begin{tabular}{|c|c|c|c|c|c|c|c|c|c|c|c|c|c|c|c|c|c|c|}
\hline \multicolumn{1}{c}{Simulations} &
\multicolumn{2}{c}{$M_{\rm DM}(h^{-1} 10^{14} M_{\odot})$} &
\multicolumn{2}{c}{$M_{\rm GAS}(h^{-1} 10^{13} M_{\odot})$} &
\multicolumn{2}{c}{$R_{\rm v}(h^{-1}$ kpc)} &
\multicolumn{2}{c}{$L_x (10^{44} \,{\rm erg\,s^{-1}})$} &
\multicolumn{2}{c}{$T_{\rm MW}$(keV)} &
\multicolumn{2}{c}{$T_{\rm SL}$(keV)}\\
\hline
 & svisc & lvisc & svisc & lvisc & svisc & lvisc & svisc & lvisc
 & svisc & lvisc & svisc & lvisc \\
g1    &14.5 &14.5 &17.5 &17.0 &2360 &2355 &47.1 & 21.3& 7.2&7.1& 5.8 &5.6 \\
g8    &22.6 &22.4 &19.8 &19.8 &2712 &2705 &63.1 & 32.1& 9.3&9.1& 6.2 &5.7 \\
g51   &13.0 &13.0 &11.5 &11.5 &2255 &2251 &30.8 & 17.9& 6.4&6.3& 4.6 &4.7 \\
g72   &13.5 &13.4 &12.0 &11.9 &2286 &2280 &18.3 & 14.1& 5.8&5.8& 4.0 &4.0 \\
g676  & 1.1 & 1.0 & 0.95& 0.91& 983 & 972 & 3.2 &  1.4& 1.3&1.3& 1.6 &1.5 \\
g914  & 1.2 & 1.0 & 1.07& 0.91&1023 & 971 & 4.2 &  1.7& 1.4&1.3& 1.6 &1.7 \\
g1542 & 1.1 & 1.0 & 0.95& 0.90& 982 & 967 & 3.0 &  1.4& 1.3&1.2& 1.4 &1.5 \\
g3344 & 1.1 & 1.1 & 1.00& 0.96&1002 & 993 & 2.2 &  1.4& 1.4&1.3& 1.4 &1.5 \\
g6212 & 1.1 & 1.1 & 1.00& 1.01&1000 &1006 & 3.0 &  1.5& 1.3&1.3& 1.6 &1.7 \\
\hline
\end{tabular}
\label{tab:char}
\end{center}
\end{table*}

\subsection{Schock-cloud interaction}

To verify that the low--viscosity scheme also works in more complex
hydrodynamical situations, we simulate a test problem where a strong shock
strikes a gas cloud embedded at pressure equilibrium in a lower density
environment. A recent discussion of this setup can be found in
\cite{2002ApJ...576..832P} and references therein. SPH is able to reproduce
the main features expected in this test problem reasonably well, like reverse
and reflected shocks, back-flow, primary and secondary Mach stems, primary and
secondary vortices, etc.~\citep[see][]{2005astro.ph..5010S}. Our purpose here
is to check whether the new scheme for a time-variable viscosity performs at
least equally well as the original approach.

In Figure \ref{fig:tube2d}, we compare the time evolution of the projected gas
density for the original viscosity scheme (left hand side) with the new
low--viscosity scheme (right hand side). Overall, we find that the new scheme
produces quite similar results as the original method. But there are also a
number of details where the low--viscosity scheme appears to work better.
One is the external reverse bow shock which is resolved more sharply with the
new scheme compared to the original one.  This is consistent with our findings
from the previous tests, where we could also notice that shocks tend to be
resolved somewhat sharper using the new scheme. We also note that
instabilities along shear flows (e.g. the forming vortexes or the back-flow)
are appearing at an earlier time, as expected if the viscosity of the
numerical scheme is lower. This should help to resolve turbulence better.

In summary, the low--viscosity scheme appears to work without
problems even in complex situations involving multiple shocks and
vorticity generation, while it is still able to keep the advantage of
a reduced viscosity in regions away from shocks. We can therefore
expect this scheme to also work well in a proper environment of
cosmological structure formation, and simulations should be able to
benefit from the reduced viscosity characteristics of the scheme.


\section{Cosmological Cluster Simulations} \label{sec:simulations}

We have performed high-resolution hydrodynamical simulations of the
formation of 9 galaxy clusters. The clusters span a mass-range from
$10^{14}\,h^{-1}{\rm M}_{\odot}$ to $2.3\times 10^{15}h^{-1}{\rm
M}_{\odot}$ and have originally been selected from a DM--only
simulation \citep{YO01.1} with box-size $479\,h^{-1}$Mpc of a flat $\Lambda$CDM model
with $\Omega_0=0.3$, $h=0.7$, $\sigma_8=0.9$ and $\Omega_{\rm
b}=0.04$.  Using the `Zoomed Initial Conditions' (ZIC) technique
\citep{1997MNRAS.286..865T}, we then re-simulated the clusters with higher mass and
force resolution by populating their Lagrangian regions in the initial
conditions with more particles, adding additional small-scale power
appropriately.  The selection of the initial region was carried out
with an iterative process, involving several low resolution DM-only
resimulations to optimise the simulated volume. The iterative cleaning
process ensured that all of our clusters are free from contaminating
boundary effects out to at least 3 - 5 virial radii.  Gas was
introduced in the high--resolution region by splitting each parent
particle into a gas and a DM particle. The final mass--resolution of
these simulations was $m_{\rm DM}=1.13\times 10^9\,h^{-1}{\rm
M}_\odot$ and $m_{\rm gas}=1.7\times 10^8\,h^{-1}{\rm M}_\odot$ for
dark matter and gas within the high--resolution region,
respectively. The clusters were hence resolved with between
$2\times10^5$ and $4\times10^6$ particles, depending on their final
mass. For details on their properties see Table \ref{tab:char}.  The
gravitational softening length was $\epsilon=5.0\, h^{-1}$kpc
(Plummer--equivalent), kept fixed in physical units at low redshift
and switched to constant comoving softening of $\epsilon=30.0\,
h^{-1}$kpc at $z\ge 5$. Additionally we re-simulated one of the
smaller cluster ({\it g676}) with 6 times more particles (HR),
decreasing the softening by a factor of two to $\epsilon=2.5\,
h^{-1}$kpc.

We computed three sets of simulations using non radiative gas dynamics, where each
cluster was simulated three times with different prescriptions for the
artificial viscosity. In our first set, we used the original formulation of
artificial viscosity within SPH. In the second set, we used the
parametrisation based on signal velocity, but with a fixed coefficient for the
viscosity. Finally, in our third set, we employed the time dependent viscosity
scheme, which we expect to lead to lower residual numerical viscosity.  Our
simulations were all carried out with an extended version of {\small GADGET-2}
\citep{2005astro.ph..5010S}, a new version of the parallel TreeSPH simulation code {\small
  GADGET} \citep{SP01.1}. We note that the formulation of SPH used in this
code follows the `entropy-conserving' method proposed by
\citet{2002MNRAS.333..649S}.


\section{Identifying Turbulence} \label{sec:turbulence}

\begin{figure}
\includegraphics[width=0.5\textwidth]{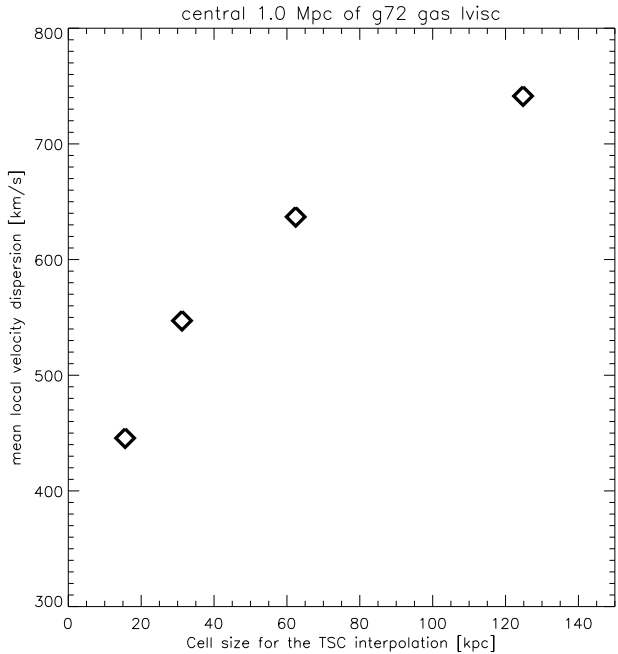}
\caption{Mean local velocity dispersion for the central
$500^{3}{\rm kpc}^{3}$ box as a function of the resolution adopted for the 
TSC--smoothing of the local mean field. Results are plotted for a low 
viscosity simulation.}
\label{fig:disp_resol}
\end{figure}

In the idealized case of homegeneus and isotropic turbulence, the
autocorrelation function of the velocity field of the fluid should not depend
on the position (homogeneity) and it should only depend on the magnitude of
the distance $\vec{r}$ between points (isotropy).  The tensor of the
correlation function of the velocities is thus given by \citep[e.g.][]{1998pfp..conf.....C}:
\begin{equation}
R_{ij}(r) = \left<
v_{i}(\vec{x}) v_{j}(\vec{x}+\vec{r})
\right> 
\label{rij}
\end{equation}
\noindent
where $\vec{x}$ is the position of a fluid particle.
The 3D power spectral density of the
turbulent field is given by \citep[e.g.][]{1998pfp..conf.....C}:
\begin{equation}
\Phi_{ij}(\vec{k})={1\over{(2\pi)^3}}\int R_{ij}(\vec {r})
\exp ( {\it i}\, \vec{k}\vec{r} )\, {\rm d}\vec{r}.
\label{Phi_ij}
\end{equation}

\noindent
The energy spectrum, $E(k)$, associated with the fluctuations of the velocity
field is related to the diagonal parts of both the tensor of the correlation
function, and that of the power spectral density.
This energy spectrum is given by \citep[e.g.][]{1998pfp..conf.....C}:
\begin{equation}
E(k) = 2 \pi k^2 \Phi_{ii}(k),
\label{ek}
\end{equation}
\noindent
and the total turbulent energy per unit mass is 
\begin{equation}
u_{\rm turb}={1\over 2} \left< v^2 \right> =
{1\over 2} R_{ii}(\vec{r}=0)=
\int_0^{\infty}
E(k)\,{\rm d}k,
\label{energy_ek}
\end{equation}
where the summation convention over equal indices is adopted.

The real case of the intracluster medium is however much more complex, in
particular not homogeneous and isotropic.  The gravitational field induces
density and temperature gradients in the ICM, and the continuous infall of
substructures drives bulk motions through the ICM. These effects break both
homogeneity and isotropy at some level, at least on the scale of the cluster,
and thus demand a more complicated formalism to appropriately characterise the
turbulent field.  It is not the aim of the present paper to solve this problem
completely. Instead we focus on a zero--order description of the energy stored
in turbulence in the simulated boxes, and for this purpose the basic formalism
described below should be sufficient.

\begin{figure*}
\includegraphics[width=0.49\textwidth]{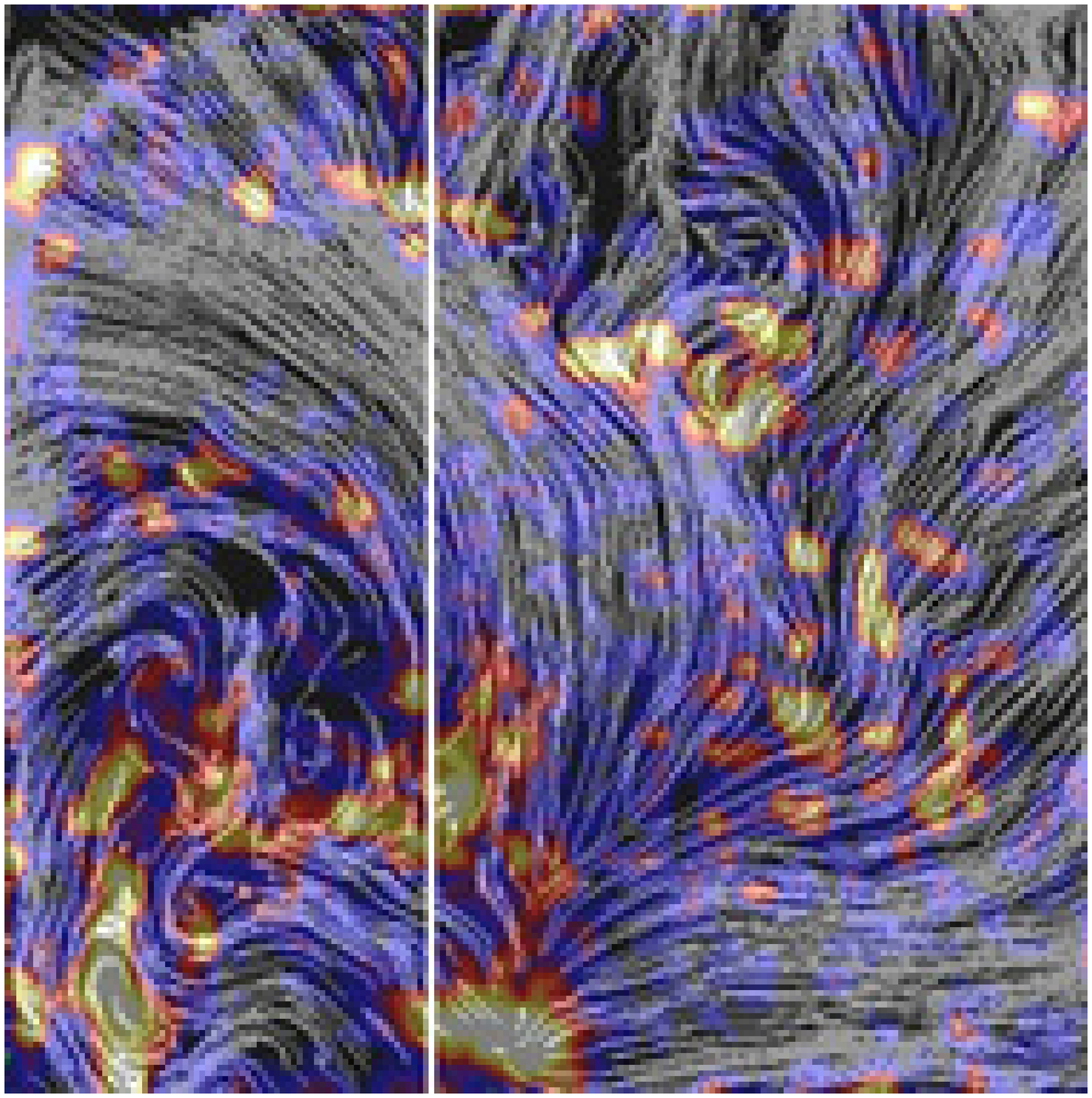}
\includegraphics[width=0.49\textwidth]{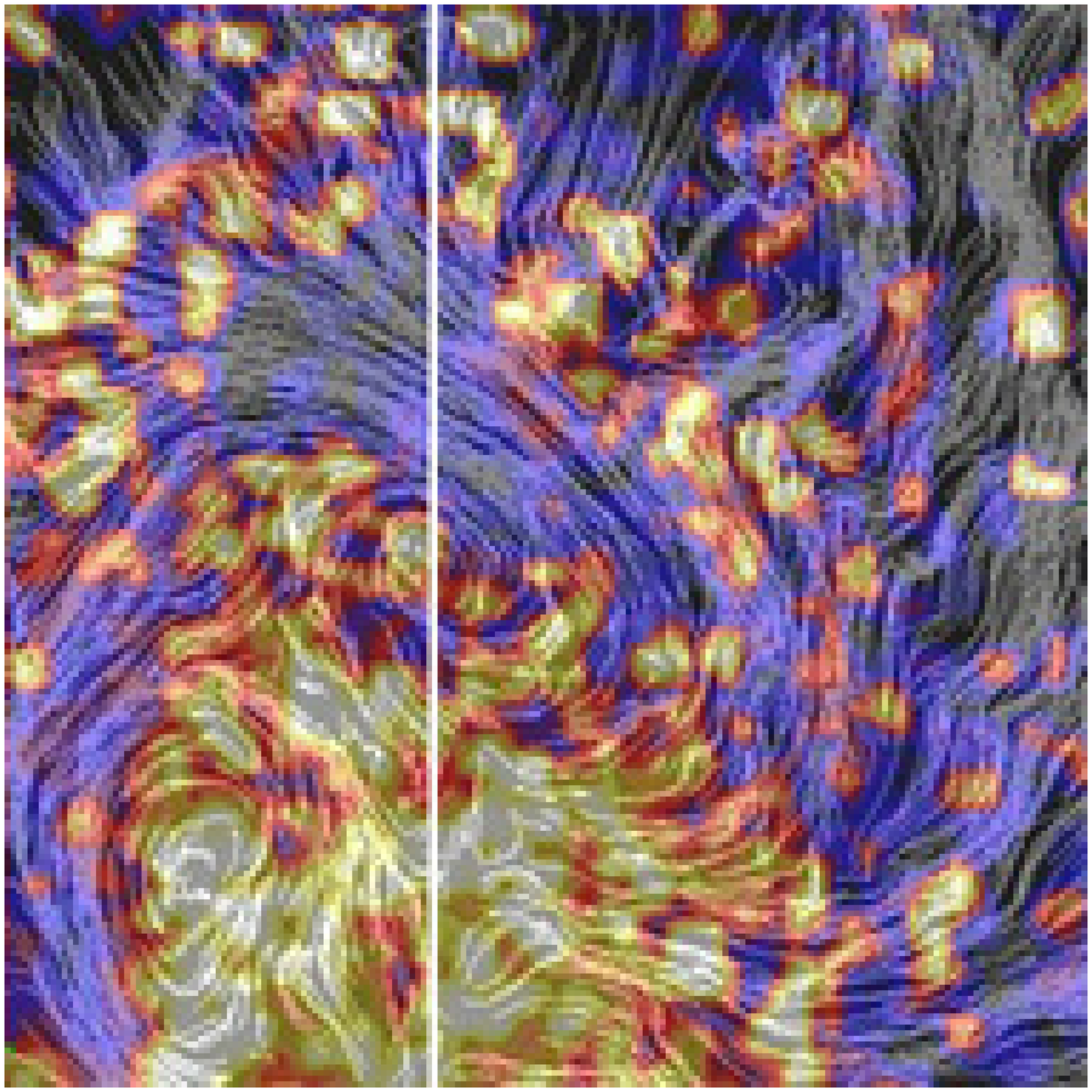}\\
\includegraphics[width=0.49\textwidth]{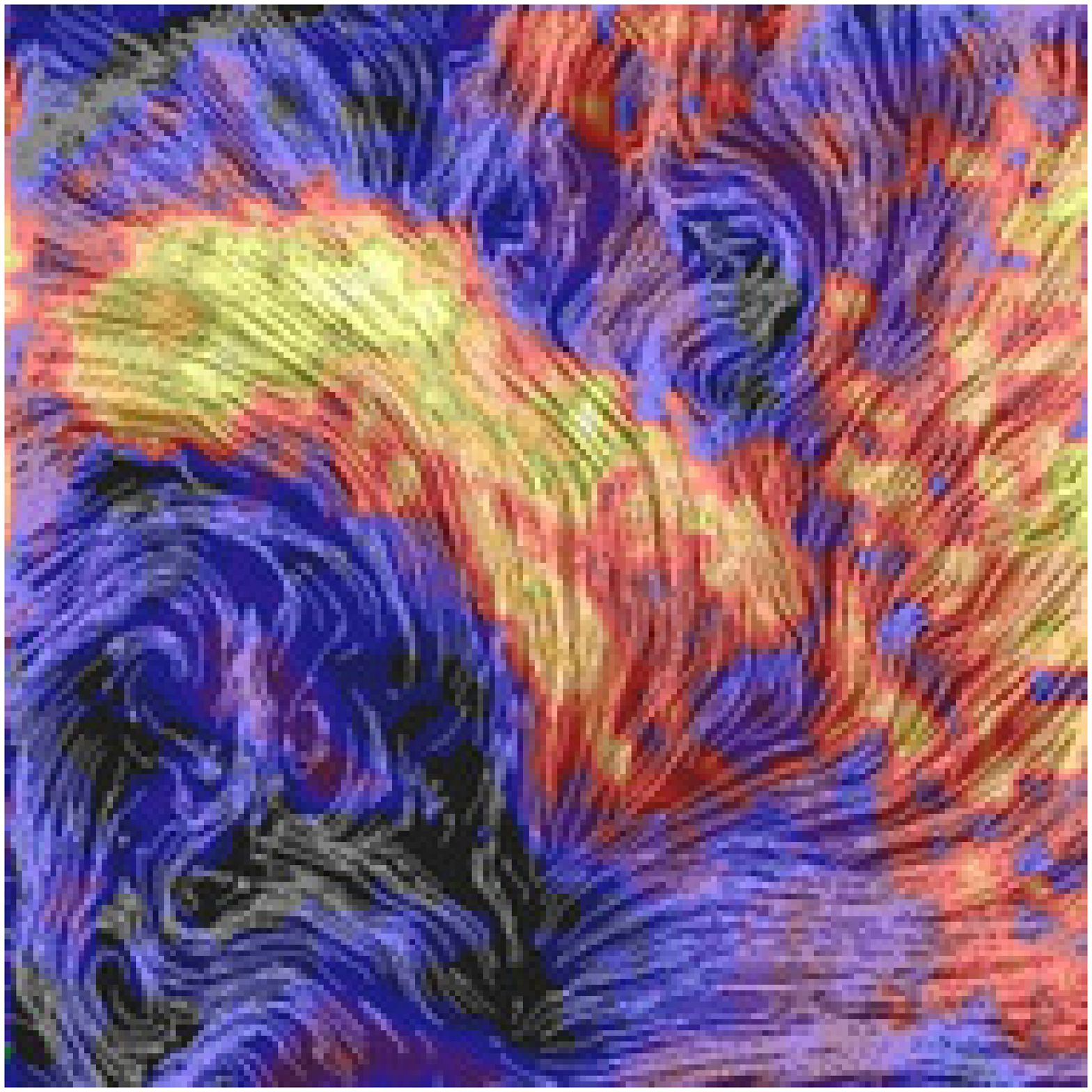}
\includegraphics[width=0.49\textwidth]{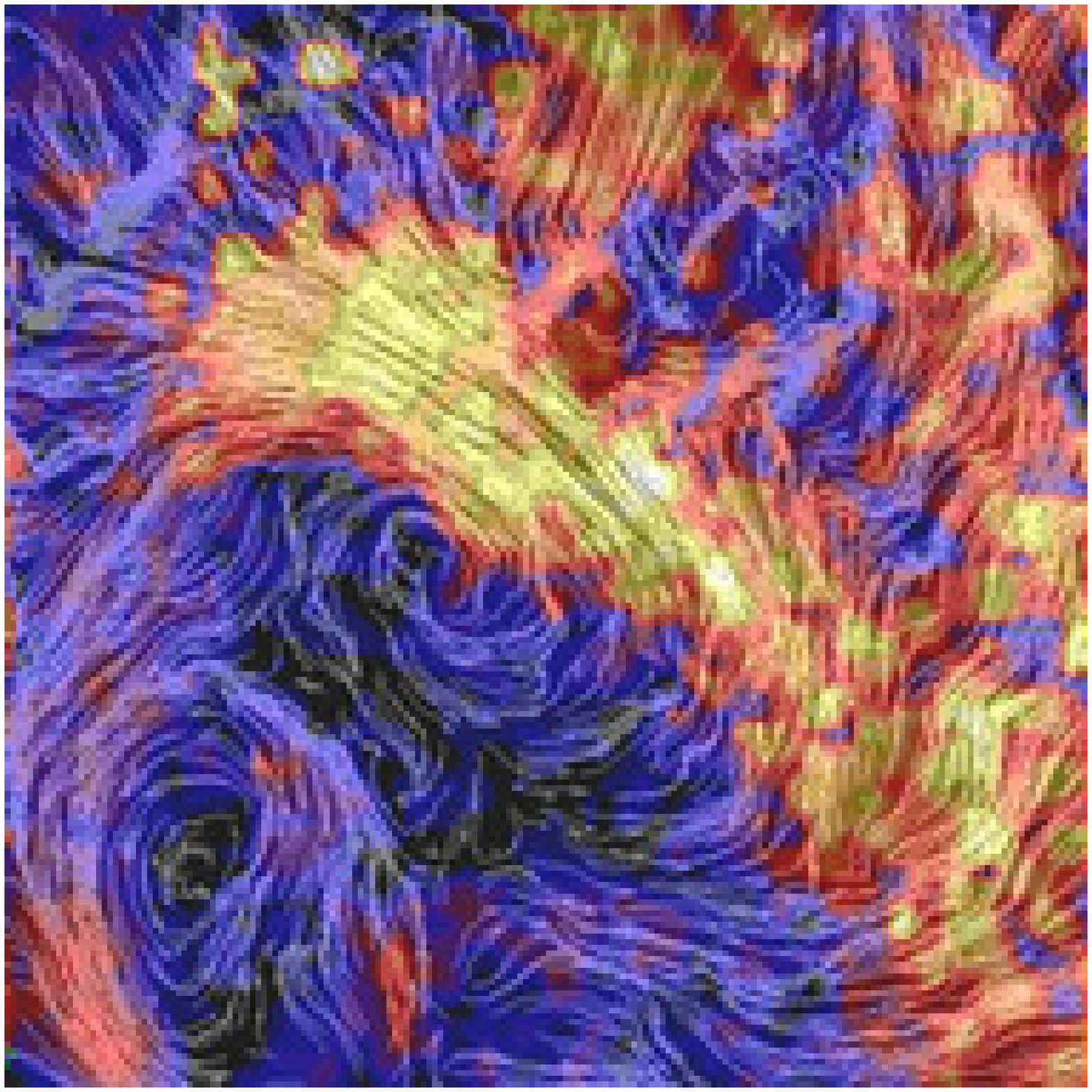}
\caption{Gas velocity field in a slice through the
central Mpc of a cluster simulation {\it g72} after subtracting the 
{\em global} mean bulk velocity of the cluster. The panels on the left is for a
run with the original viscosity of SPH while the panels on the right shows
the result for the low viscosity scheme. The underlying colour maps
represent the turbulent kinetic energy content of particles, inferred
using the local velocity method (upper row) or the standard velocity
method (lower row). For the local velocity method a 
conservative $64^3$ grid is used in the TSC smoothing. 
The cluster centre is just below the lower-left corner of the images. The
vertical lines in the upper row show where the 1--dimensional profile for the simulated
radio--emission of Fig.~\ref{fig:radio1d} are taken.}
\label{fig:vel_mean}
\end{figure*}

\begin{figure*}
\includegraphics[width=0.49\textwidth]{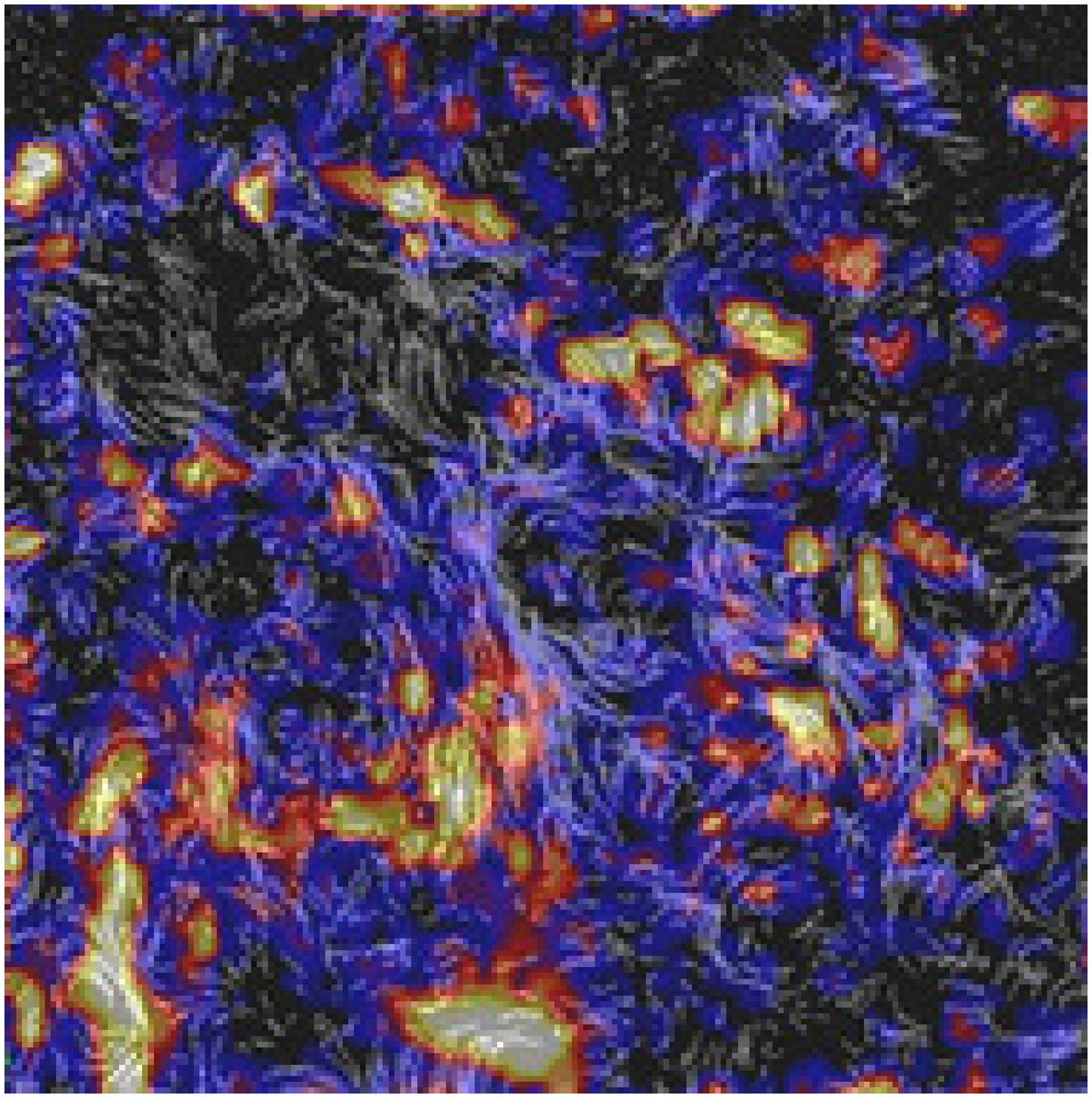}
\includegraphics[width=0.49\textwidth]{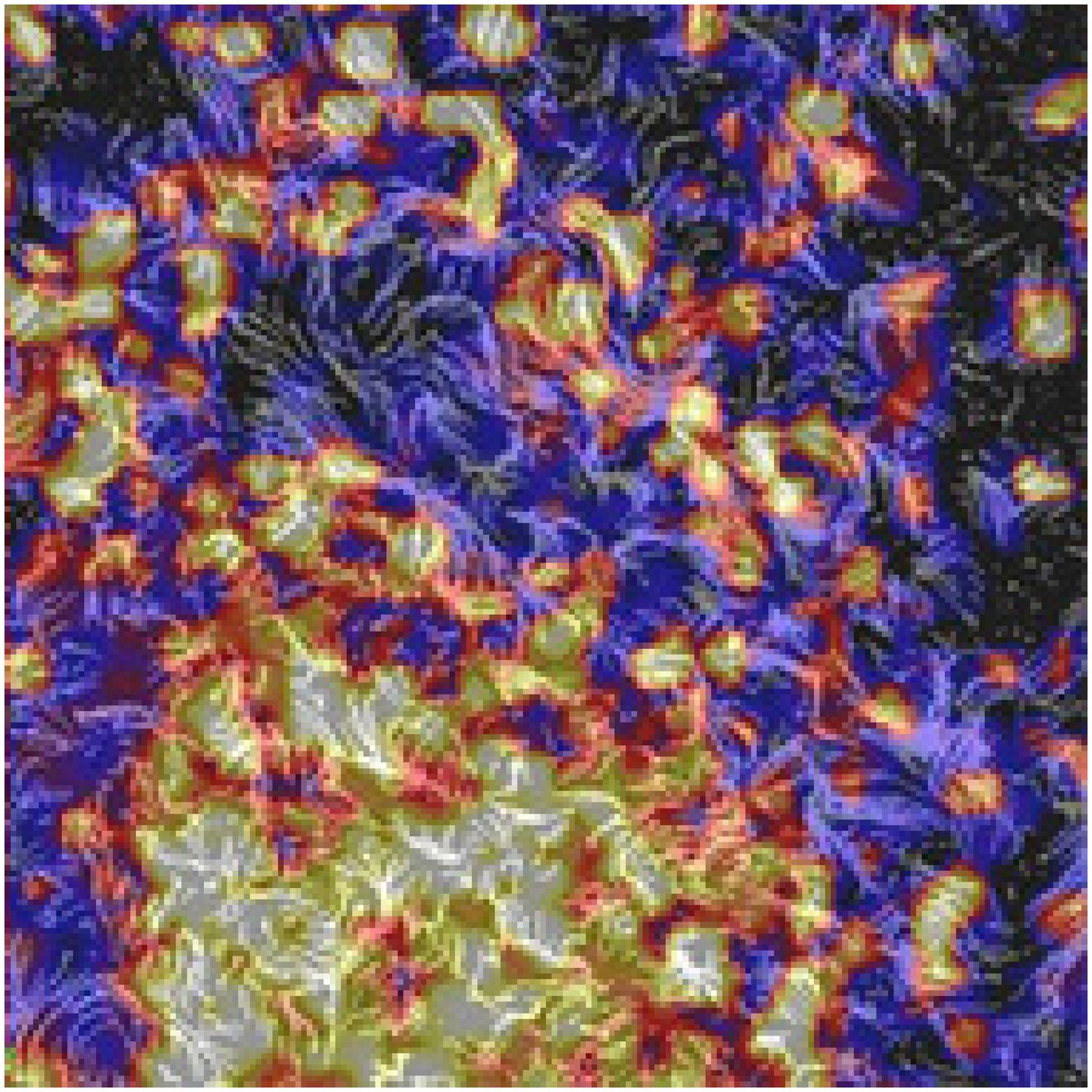}
\caption{Same slice of the Gas velocity field as in figure \ref{fig:vel_mean}
  of cluster {\it g72} after subtracting the {\em local} mean velocity of the cluster.
  The panel on the left is for a run with the original viscosity of SPH while
  the panel on the right shows the result for the low viscosity
  scheme.}
  \label{fig:vel_local}
\end{figure*}

A crucial issue in describing turbulent fields in the ICM is the distinction
between large-scale coherent velocity field and small-scale `random' motions.
Unfortunately, the definition of a suitable mean velocity field is not
unambiguous because the involved scale of averaging introduces a certain
degree of arbitrariness.  Perhaps the simplest possible procedure is to take
the mean velocity computed for the cluster volume (calculated, for example,
within a sphere of radius $R_{\rm vir}$) as the coherent velocity field, and
then to define the turbulent velocity component as a residual to this
velocity. This simple approach (hereafter {\it standard} approach) has been
widely employed in previous works
\citep[e.g.,][]{1999rgm87conf..106N,2003AstL...29..783S}, and 
led to the identification of ICM turbulence in these studies.  However, an obvious
problem with this method is that this global subtraction 
fail to distinguish a pure laminar bulk flow from a turbulent pattern of motion.  Note
that such a large scale laminar flows are quite common in cosmological
simulations, where the growth of clusters causes frequent infalls and
accretions of sub--halos.  This infall of substructures is presumably one of
the primary injection mechanisms of ICM turbulence.

To avoid this problem, a mean velocity field smoothed on scales smaller than
the whole box can be used, and then the field of velocity fluctuations is
defined by subtracting this mean--local velocity, $\overline{\vec{v}}(\vec{x})$, from the
individual velocities $\vec{v}_{i}$ of each gas particle.  We note that if the
smoothing scale is chosen too small, one may risk loosing large eddies in the
system if they are present, but at least this procedure does not overestimate
the level of turbulence.

Following this second approach (hereafter {\it local--velocity} approach), we
construct a mean local velocity field $\overline{{\vec{v}}}(x)$ on a uniform
mesh by assigning the individual particles to a mesh with a {\it Triangular
  Shape Cloud} (TSC) window function.  The mesh covers a region of 1.0
comoving Mpc on a side and typically has between $8^3$ and $64^3$ cells, which
is coarse enough to avoid undersampling effects. The equivalent width of the
TSC kernel is approximatively 3 grid cells in each dimension, corresponding to
a smoothing scale of $\approx 360-45$ kpc, respectively.  As our analysis is
restricted only to the highest density region in the clusters, the scale for
the TSC--smoothing is always larger than the SPH smoothing lengths for the gas
particles, which typically span the range $7.5 - 15\, h^{-1}{\rm kpc}$ in the
box we consider.

We then evaluate the local velocity dispersion at the position
$\vec x$ of each mesh cell over all particles $a$ in the
cell by:
\begin{equation}
\sigma^2_{ij}(\vec x) \simeq \left< \left[v_{a,i} -
\bar{v}_i(\vec x)\right]\left[v_{a,j} - \bar{v}_j(\vec x)\right]
\right>_{\rm cell}, \label{sigmai}
\end{equation}
where $i$ and $j$ are the indices for the three spatial coordinates, and
$\langle \rangle_{\rm cell}$ denotes the average over particles within each
cell.

\noindent
The diagonal part of the tensor of the correlation function
of the field of velocity fluctuations at $r=0$ in the simulated box can  then 
be approximated by
\begin{equation}
R_{ii}({r}=0)\simeq \left< \sigma^2_{ii}(\vec{x})
\right>_{\rm Box} .
\label{rii}
\end{equation}

\noindent
Based on Equation (\ref{energy_ek}), we can then 
estimate
the energy density of the turbulence in real space
as
\begin{equation}
\rho(\vec{x}) \int E(k) \, {\rm d}k \sim 
{1\over 2} \rho(\vec{x}) 
\times \left\lbrace \begin{array}{lll}
\left< \sigma^2_{ii}(\vec{x})
\right>_{\rm box} \, , \\
 \\
\left< v^2_{i}(\vec{x})
\right>_{\rm box} \, ,
\end{array}
\right.
\label{ek_box}
\end{equation}
\noindent
in the local--velocity and standard case, respectively. 
Here $\rho(\vec{x})$ is the gas density within the cells.

The subtraction of a local velocity from the velocity distribution of the
particles is expected to efficiently filter out the contribution from laminar
bulk--flows with a scale $\geq$ 3 times the size of the cells used in the TSC
smoothing.  However, a large-scale turbulent velocity field component, if it
exists, would also be suppressed, so that this procedure can be expected to
reduce the turbulent velocity field to a certain degree.  As shown in
Figure~\ref{fig:disp_resol}, this depends on the resolution of the mesh used
in the TSC assignment.  Fig.~\ref{fig:disp_resol} shows that the increase of
the turbulent velocity dispersion with the cell size is not dramatic for cell
sizes larger than 100 kpc. We find that (Vazza et al., in
prep.) a TSC smoothing with larger cell sizes would
not efficiently filter out contributions from laminar bulk--motions. It can be
tentatively concluded that the local velocity approach with a smoothing with
$16^3-32^3$ cells in the central $(1.0\,{\rm Mpc})^{3}$ volume catches the
bulk of the turbulent velocity field in the simulated box.  Therefore, if not
specified otherwise, all the numerical quantities given in the following are
obtained using a TSC--assignment procedure based on $32^3$ cells.  A more
detailed discussion of this method and tests of the parameters involved is
reported elsewhere (Vazza et al., in prep.).

Figures \ref{fig:vel_mean} and \ref{fig:vel_local} give examples of the
turbulent velocity field calculated with both the standard and local velocity
methods, showing the same galaxy cluster in both cases, but in one case
simulated with the signal--velocity variant of the viscosity, and in
the other with the new time-dependent low--viscosity scheme. Note that we here
selected a situation where a large (ca. 500 kpc long) laminar flow pattern can
be easily identified close to the centre of one of our simulated clusters
({\it g72}). When the mean cluster velocity field is subtracted as in
Figure~\ref{fig:vel_mean}, large residual bulk flow patterns remain visible,
caused by a substructure moving through the cluster atmosphere. We
colour-coded the turbulent kinetic energy of particles, $E_t(\vec{x}) \sim
1/2\, \rho(\vec{x}) \sigma_v(\vec{x})^2$, after subtracting the local mean
velocity field (here smoothed onto a $64^{3}$ mesh) for the upper
panels and aster subtracting the {\em global} mean bulk velocity of
the cluster for the lower panels. One can see that fluid
instabilities of Kelvin-Helmholtz type are growing along the interfaces of the
large laminar flow pattern, visible in the upper left panel.  As expected, the
strength of this turbulent velocity field is considerably larger in the
simulation obtained with the new low--viscosity scheme, providing evidence that
such instabilities are less strongly damped in this scheme. This can also be
seen by the longer flow field lines in
Figure~\ref{fig:vel_local}. Figures \ref{fig:vel_mean} also visually
confirms the differences in the two approaches of filtering the
velocity field. Whereas the local--velocity approach highlights the
energy within the velocity structure along boundary layers, the energy
within the large, bulk motions are preferentially selected when only
subtraction the {\em global} mean bulk velocity.  

The total cumulative kinetic energy in the random gas motions inside our mesh
(cantered on the cluster centre) reaches 5\%-30\% of the thermal energy for
the simulations using the new, low--viscosity scheme, whereas it stays at much
lower levels ($\approx$2\%-10\%) when the signal velocity parameterisation of
the viscosity is used. If the original viscosity scheme is used, it is
typically at even lower values ($\approx$1\%-5\%).

In general, we find that more massive clusters tend to have a higher fraction
of turbulent energy content. However, given that our simulations have fixed
mass resolution, this trend could in principle simply reflect a numerical
resolution effect. In order to get further information on this, we have
re-simulated one of the smaller clusters ({\it g676}) with 6 times better mass
resolution using the signal velocity parameterisation of the viscosity.  At
$z=0$, this cluster is then resolved by nearly as many particles as the
massive clusters simulated with our standard resolution. We find that for this
high-resolution simulation the level of turbulence ($\approx 3$\%) 
is increased compared with the normal resolution ($\approx 2$\%), 
but it stays less to what we found for the
low--viscosity scheme at our normal resolution ($\approx 5$\%).        
This confirms two
expectations. First, the low viscosity scheme effectively increases the
resolution on which SPH simulations can resolve small-scale velocity
structure, which otherwise gets already suppressed on scales larger than the
smoothing length by spurious viscous damping effects due to the artificial
viscosity. Second, the amount of turbulence in the high resolution version of
{\it g676} is still less than what we find with the same viscosity
implementation in the larger systems, and even much smaller than what we find
with the low--viscosity scheme in the large clusters. This tentatively suggests
that the trend of a mass-dependence of the importance of turbulence is not
caused by numerical effects. Note that with a fixed physical
scale of $1\,{\rm Mpc}$ we are sampling different fractions of $R_{\rm vir}$
in clusters of different masses. However, if, in case of the less massive
system, we restrict the sampling relative to $R_{\rm vir}$ to measure
within comparable volumes, the fraction of turbulent energy content  
found in the small cluster increase roughly by a factor of
two. Thereby we still find a significant trend with mass when
measuring turbulence within a fixed fraction of  $R_{\rm
vir}$. Although it should be mentioned, that unless the dissipation of
turbulence on small scales will me modeled correctly in a physical
granted way, the different formation time scales of systems with
different masses can potentially also contribute to such a trend.

\begin{figure}
\includegraphics[width=0.5\textwidth]{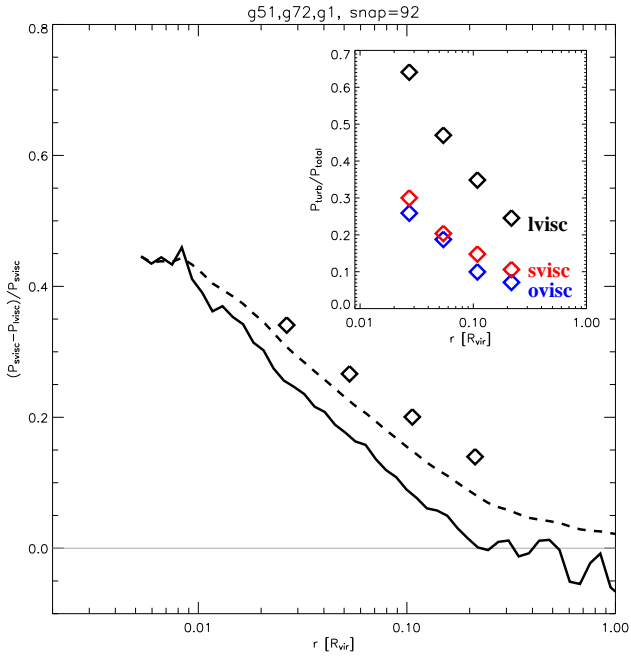}
\caption{Radial profile of the relative thermal pressure
  difference averaged over three nearly equally massive clusters ({\it
    g1},{\it g51} and {\it g72}), comparing the signal velocity based
and low--viscosity runs (lines). The dashed line is the
  cumulative difference, whereas the solid line marks the profile in radial
  shells. The diamonds mark the difference in the turbulent energy support we
  inferred from the local velocity dispersion within several concentric cubes of
  different sizes ($l_{\rm cube}=2r$) for the same runs. This should be compared
with the dashed line.  The inlay
  shows the absolute value inferred from the local velocity dispersion from
  the different viscosity parameterisations, respectively.}
\label{fig:p_prof}
\end{figure}

In order to verify that our method for measuring the local velocity dispersion
gives reasonable values, Figure~\ref{fig:p_prof} shows a radial profile of the
volume-weighted, relative difference between thermal pressure
for the signal velocity based and
low--viscosity run. Here we used the an average over the three massive clusters
({\it g1},{\it g51} and {\it g72}) which have comparable masses. The solid
line shows the relative difference in radial shells and indicates that the
turbulent pressure support can reach even up to 50\% in the central part and
drops to 0 at approximately 0.2 $R_{\rm vir}$. The dashed line shows the
cumulative difference, which over the total cluster volume contributes between
2\% and 5\% to the total pressure. The diamonds mark the measurement inferred
from the local velocity dispersion within centred boxes of various sizes.  We
also calculate the difference between the signal velocity based
and low--viscosity runs
using the mean values over the three clusters.  Qualitatively, there is good
agreement of results obtained with this approach with the cumulative curve.
This confirms that our method to infer the turbulent energy content from the
local velocity dispersion of the gas is meaningful. Note that the temperature
which is used to calculate the pressure is determined by strong shock heating.
As different resimulations of the same object can lead to small (but in this
context non-negligible) timing differences, this can introduce sizable
variations in the calculated pressure, especially during merging events. We
verified that these differences for individual clusters are significantly
larger than the differences between the cumulative curve (dashed line) and the
data points from the local velocity dispersion (diamonds).  Therefore we can
only say that the two methods agree well within their uncertainties.

Finally, the inlay of Figure~\ref{fig:p_prof} gives the absolute contribution
from the low--viscosity, the original viscosity in its two variants using the
local velocity dispersion respectively.  It seems that using the signal based
viscosity in general leads already to more turbulence than the ``old''
original viscosity, but the time-dependent treatment of the viscosity works
even more efficiently.

\begin{figure}
\includegraphics[width=0.5\textwidth]{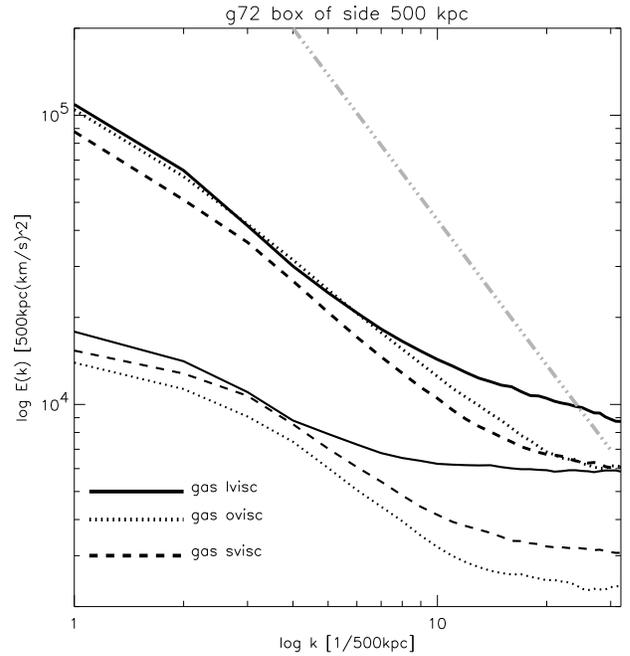}
\caption{The energy spectra of the standard velocity fluctuations ({\it
    upper curves}) and of the local velocity fluctuations ({\it lower curves})
  of gas particles in the central $500^{3}{\rm kpc}^{3}$ region of a cluster
  simulated with the original recipe for the artificial viscosity, with
  signal--velocity and with the low viscosity implementation. Additionally a
  Kolmogorov slope (dot-dot-dashed) is drawn for comparison.}
\label{fig:spect_1}
\end{figure}

Although we are using a formalism which is suitable only for isotropic and
homogeneous turbulence, the study of the turbulent energy spectrum may provide
some useful insight.  In the local mean velocity approach, we can obtain the
diagonal part of the turbulent energy spectrum using Equation~(\ref{Phi_ij}),
with $R_{ii}$ approximated as
\begin{equation}
R_{ii}(r)= \left< \left[{v}_{a,i} - 
\overline{{v}}_i(\vec{x}_a) \right] \left[{v}_{b,i} -
\overline{{v}}_i(\vec{x}_b) \right] \right>_{\rm box},
\label{rii_tsc}
\end{equation}
where $\overline{\vec{v}}(\vec{x}_a)$ is the TSC--mean velocity of the cell
which contains the point $\vec{x}_a$, and the average is over all pairs
$(a,b)$ in the box with a certain distance $r$.
\noindent
In the standard approach, we would here subtract the centre-of-mass 
velocity of the cluster instead.

A major problem for estimating the correlation functions $R_{ii}(r)$ in this
way, and with the energy spectrum calculated from SPH simulations (and in
general from adaptive resolution approaches), is given by the non--uniform
sampling of the point masses in the simulated box.  To reduce this problem we
focus on regions corresponding to the cores of galaxy clusters.
Here the requirement of isotropic and homogeneous turbulence is
hopefully better full filled. Also the 
density profile is relatively flat such that the sampling with gas particles
is not too non-uniform. In addition, we estimate the correlation function as
an average of dozens of Monte--Carlo extractions of gas particles from the
simulated output, where we picked one particle from each of the $(15.6\,{\rm
  kpc})^{3}$ cells in order to have a uniform, volume-weighted set of
particles.

In Figure~\ref{fig:spect_1}, we show examples for the energy spectra we
obtained for the two approaches by Fourier-transforming the measurements for
$R_{ii}(r)$.  The energy spectra for the two methods for treating the mean
velocities are reasonably similar in shape, but the spectrum calculated with
the local mean velocity has a lower normalisation, independent of resolution.
This is expected because the TSC smoothing filters out contributions from
laminar motions, and may also damp the turbulent field at some level.  Both
spectra show a slope nearly as steep as a Kolmogorov spectrum (which has $E(k)
\propto k^{-5/3}$) at intermediate scales, but exhibit a significant
flattening at smaller scales (i.e. large $k$). The flattening at small scales
could be caused by numerical effects inherent in the SPH technique, where an
efficient energy transfer to small-scale turbulent cells on scales approaching
the numerical resolution is prevented, and thus a complete cascade cannot
develop. Additional, the lack of numerical viscosity in the
low--viscosity scheme can in principle lead to an increase of the
noise level within the velocity field representation by the SPH
particles on scales below the smoothing length. Such noise in general could
contribute to the flattening at small scales. It is however not clear how to
separate this noise component from a turbulent cascade reaching scales
similar or below the smoothing length. Therefore one focus of future
investigations has clearly be towards this issue. 
It is however important to note that the largest turbulent energy
content (expecially at small scales) is always found in the clusters simulated
with the low--viscosity scheme.  This is particularly apparent in the
energy
spectra when the local velocity approach is used and suggests that the
energy
spectrum obtained with the standard approach is significantly affected by
laminar bulk--flows, which are not sensitive to a change in parameterisation
of the artifical viscosity.


\begin{figure*}
\includegraphics[width=0.49\textwidth]{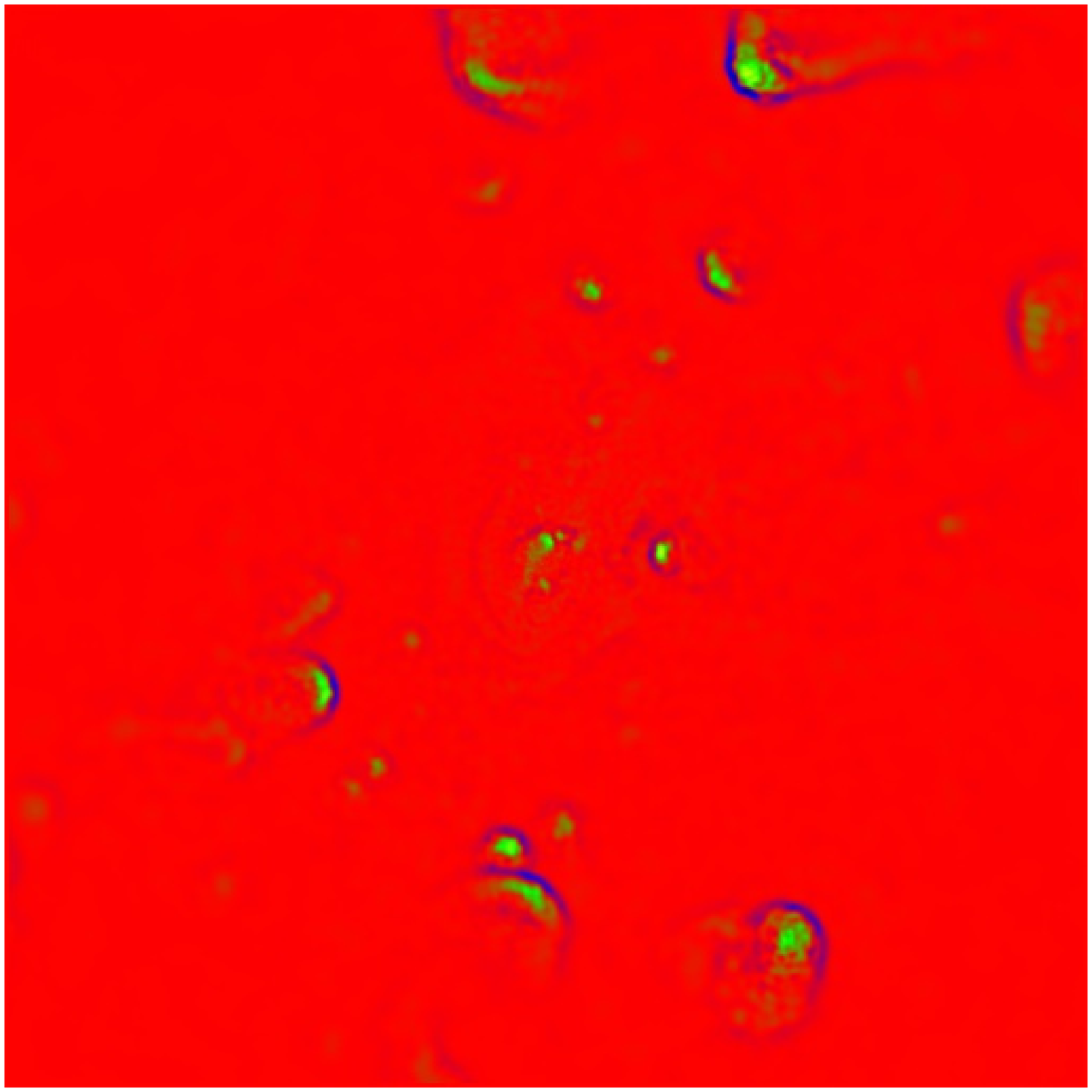}
\includegraphics[width=0.49\textwidth]{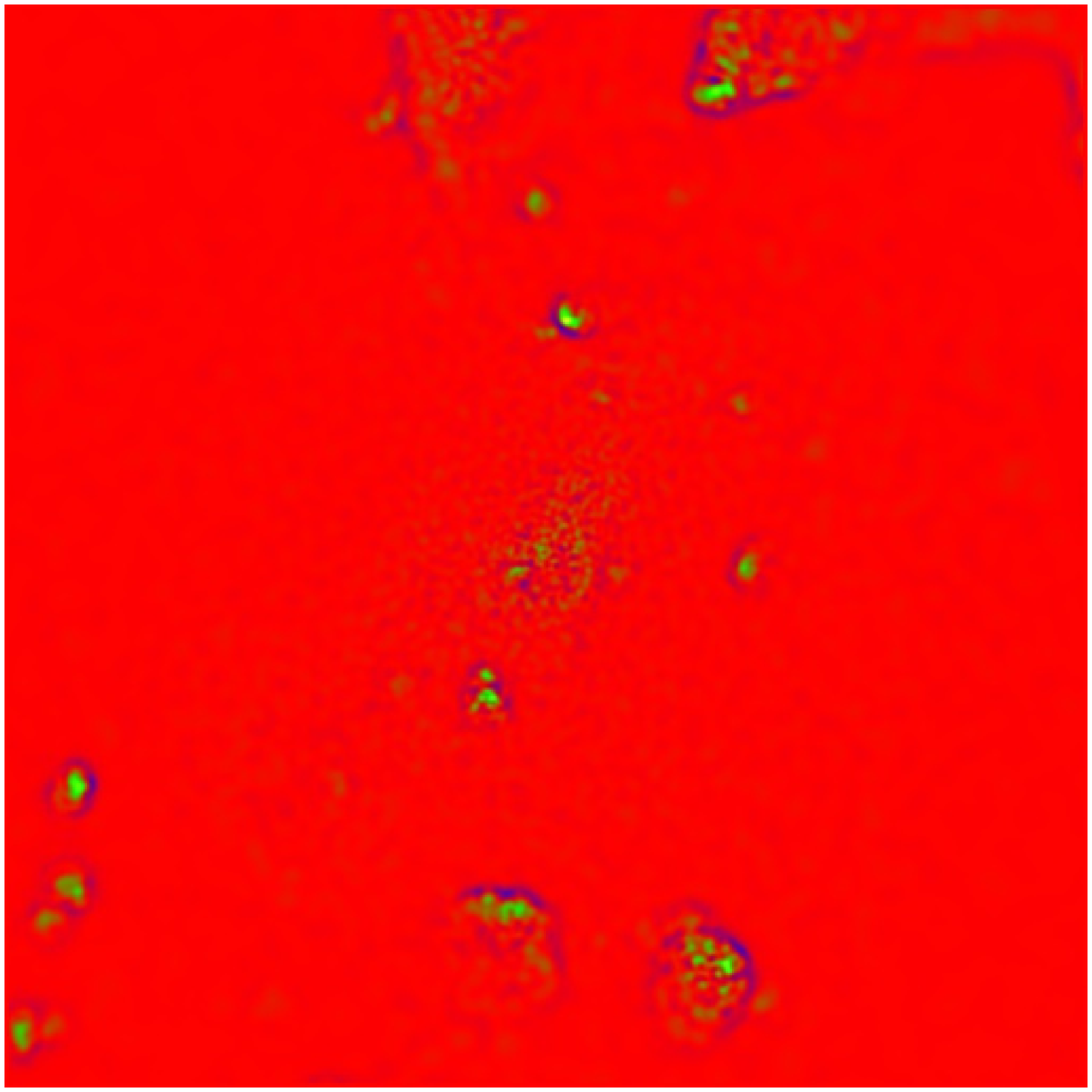}\\
\includegraphics[width=0.49\textwidth]{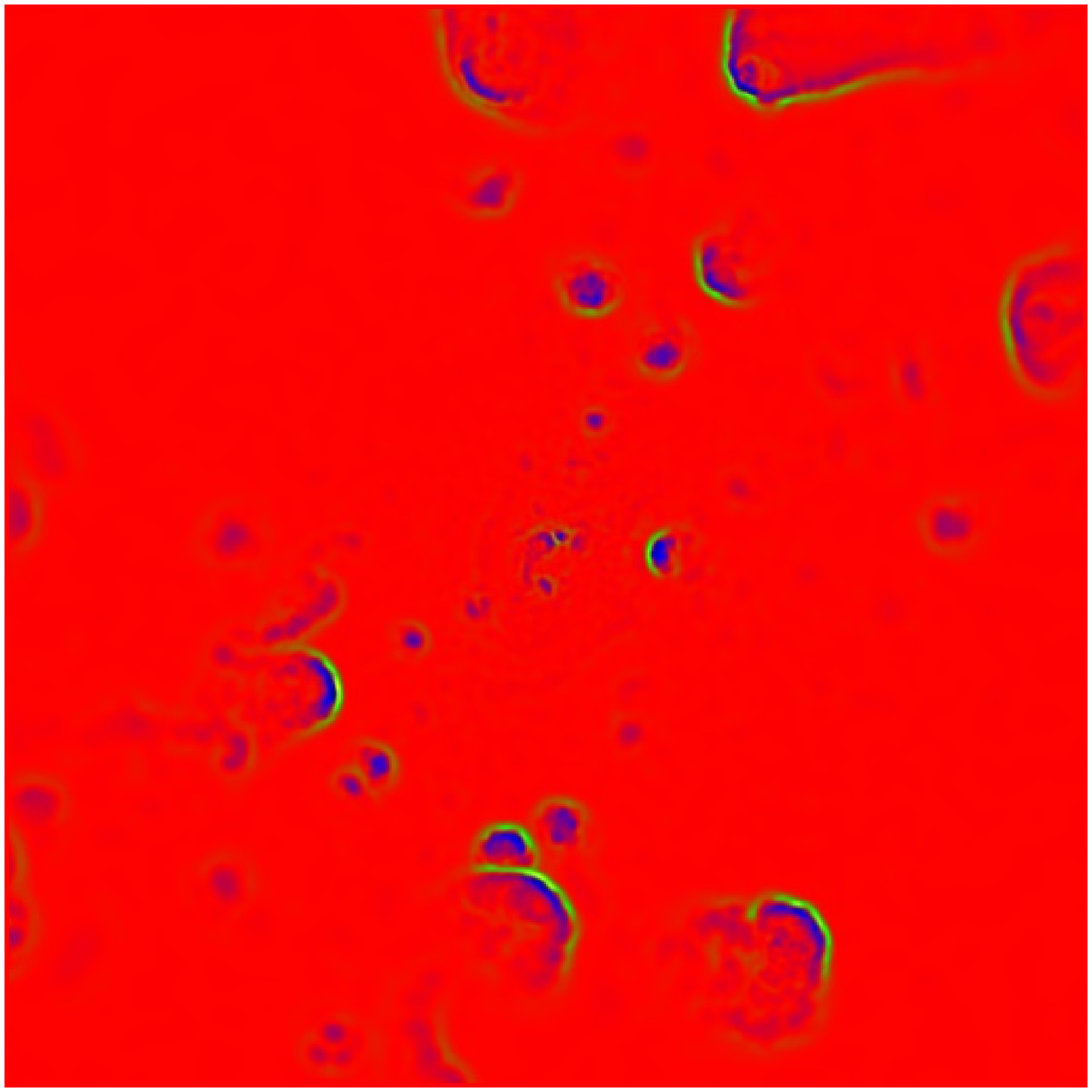}
\includegraphics[width=0.49\textwidth]{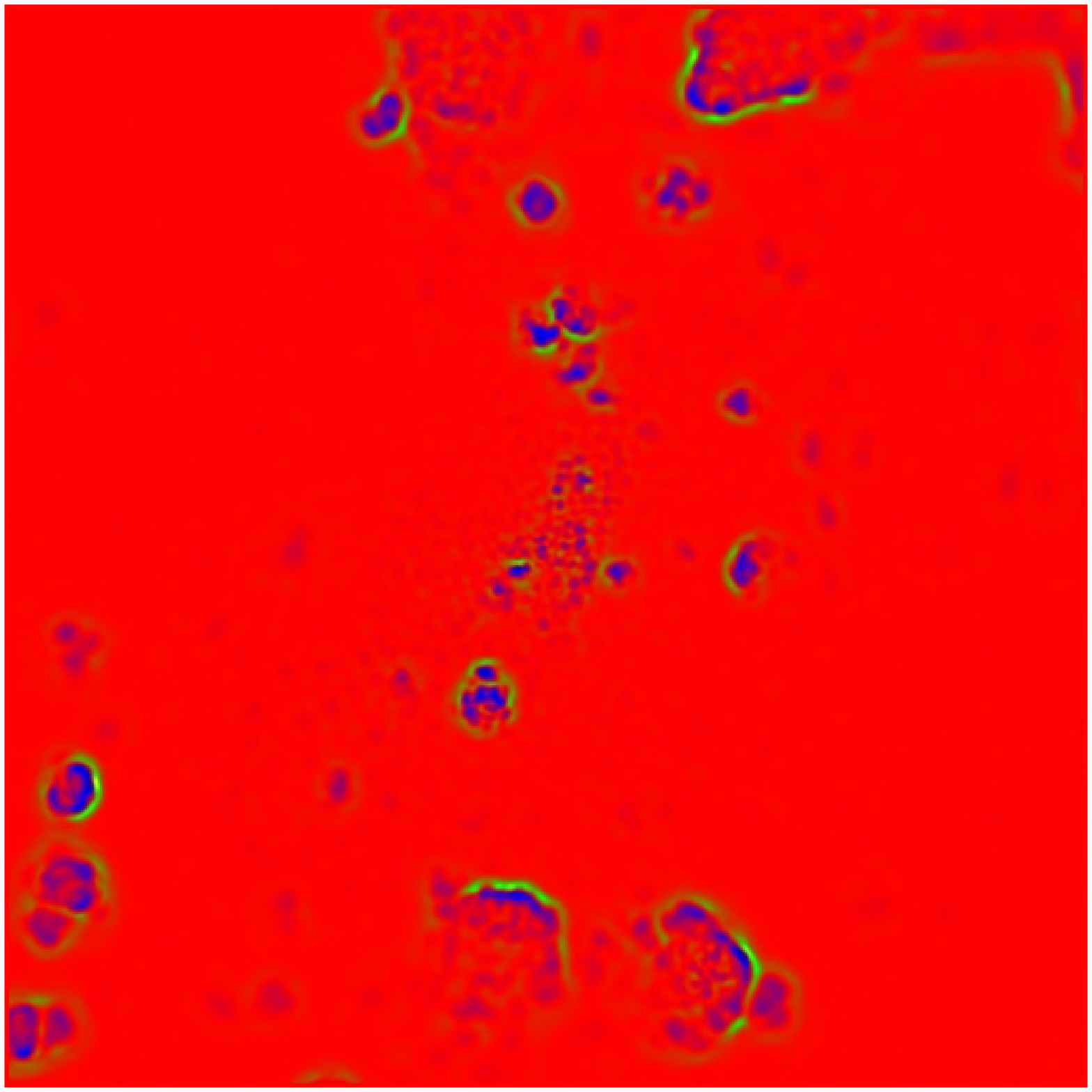}\\
\caption{ Unsharp mask images of X-ray maps for one of the massive
  clusters ({\it g1}), comparing runs with the low--viscosity scheme (right
  panels) with the original SPH scheme (left panels). The top row gives maps
  of surface brightness, while the bottom one compares maps of the
  `spectroscopic like' temperature, both within 2 Mpc centred on the cluster.
  We can see evidence for an increased level of turbulent motions behind the
  infalling substructures, and a break-up of fluid interfaces for the
  reduced viscosity scheme is clearly visible. Also, there is a general
  increase of turbulence (appearing as lumpiness) towards the centre. However,
  the most prominent signals in the map stem from the higher density or
  different temperature of substructures relative to their surrounding, or
  from shocks and contact discontinuities. For this reason, turbulence can
  be better identified in pressure maps (see Figure~\ref{fig:pmaps}). }
\label{fig:lxtxmaps}
\end{figure*}

\begin{figure*}
\includegraphics[width=0.49\textwidth]{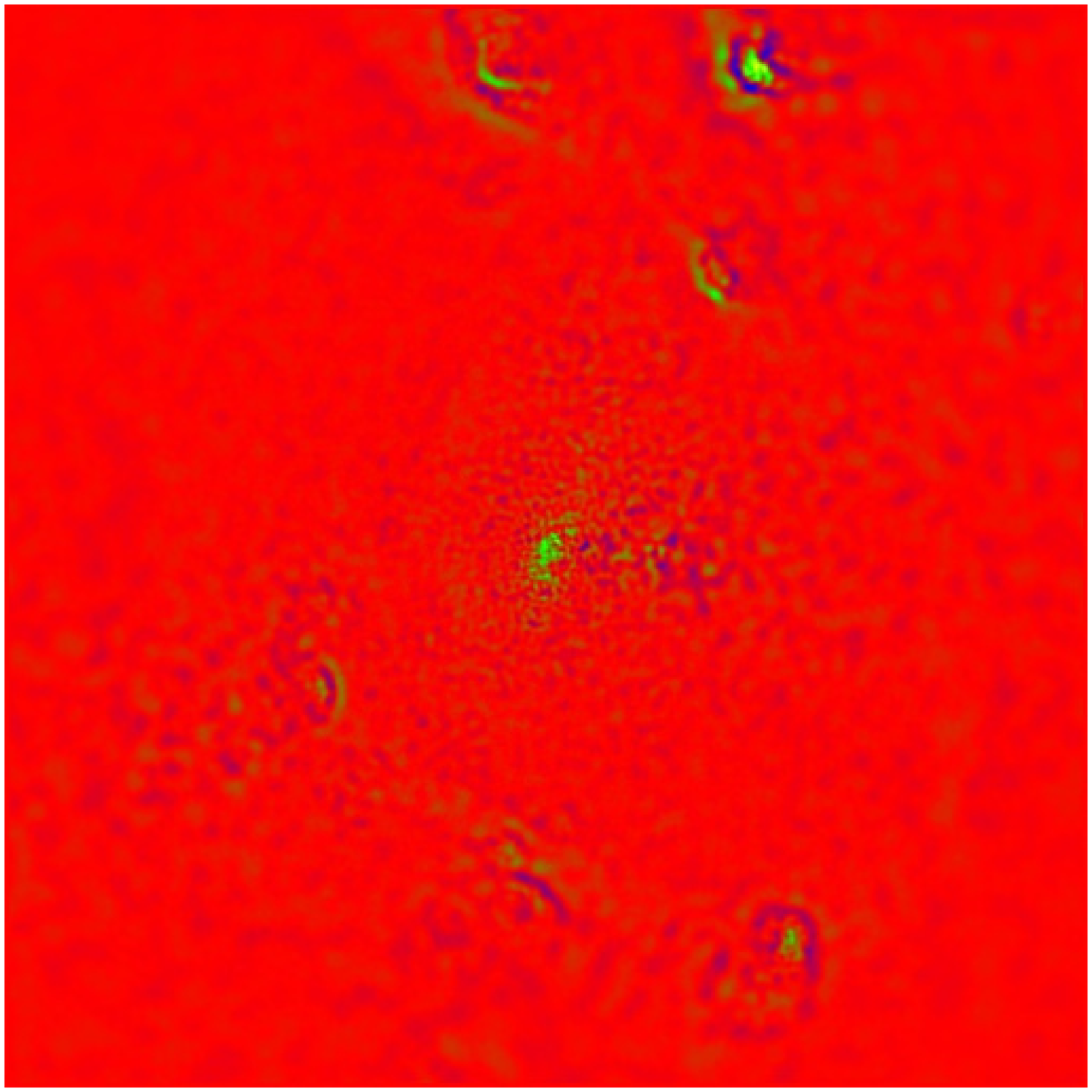}
\includegraphics[width=0.49\textwidth]{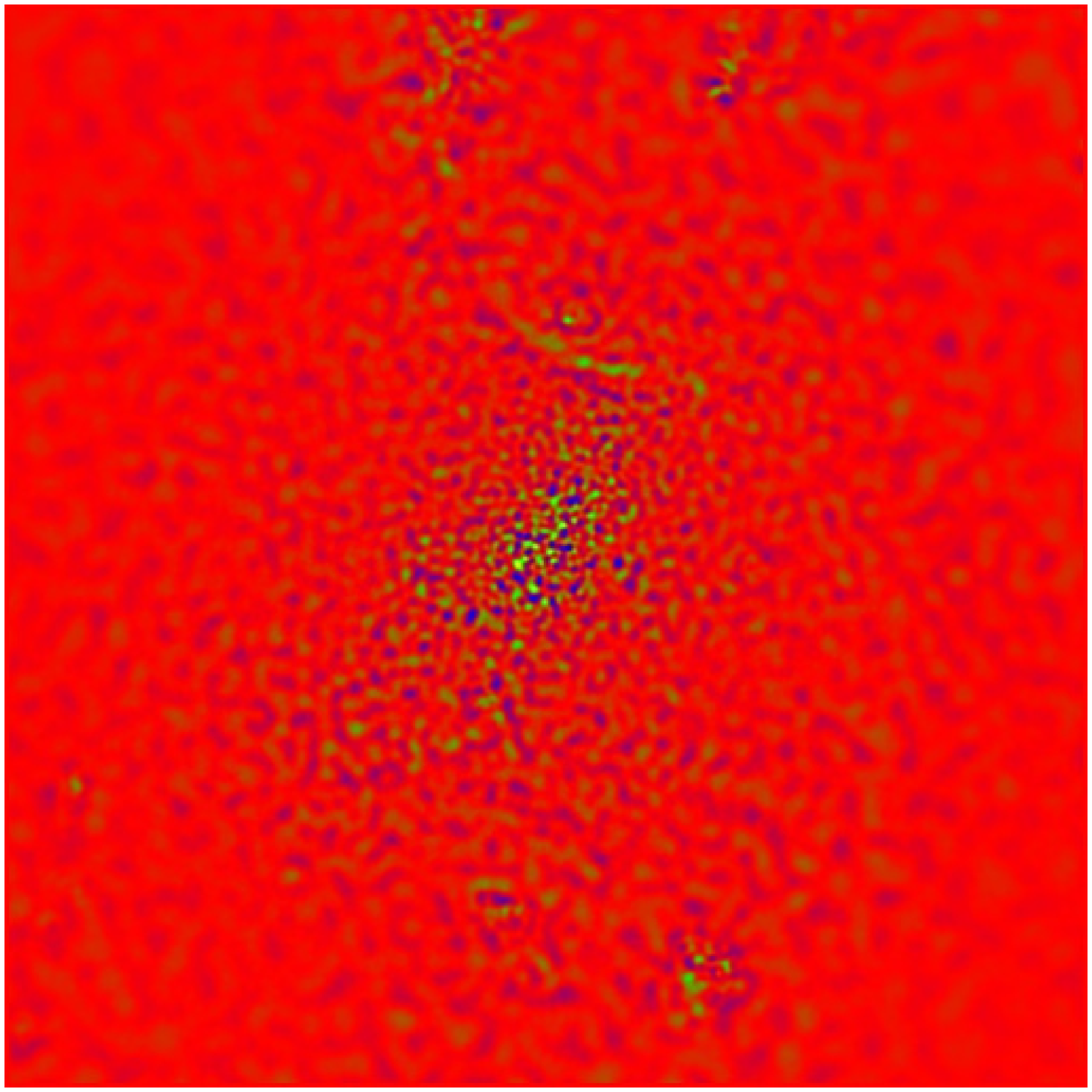}\\
\includegraphics[width=0.49\textwidth]{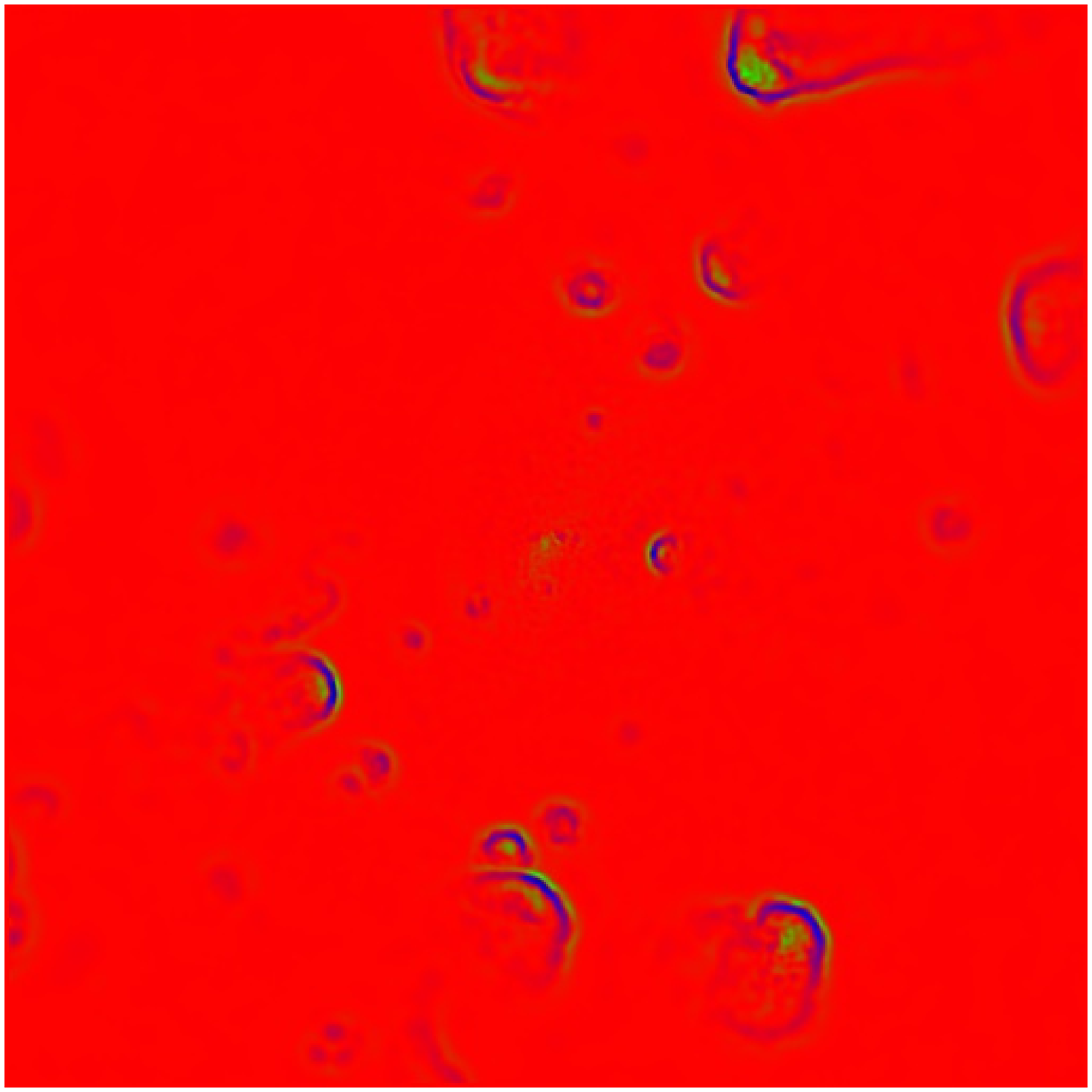}
\includegraphics[width=0.49\textwidth]{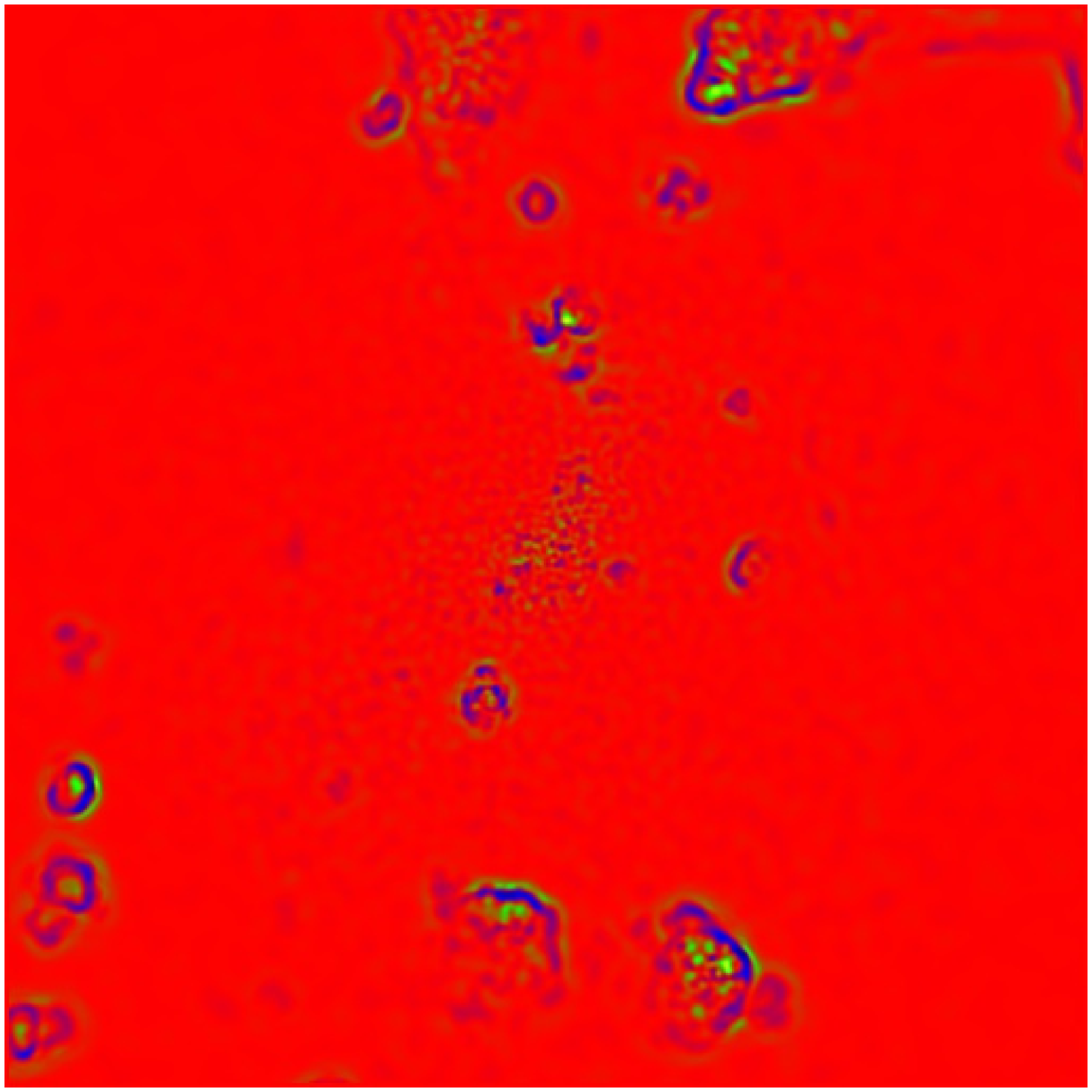}\\
\caption{Unsharp mask images of pressure maps of one of the
massive clusters ({\it g1}), comparing runs with the low--viscosity
scheme (right panels) with the original SPH scheme (left panels).  We
also compare different methods for determining the pressure maps.  The
panels of the top row show Compton-$y$ maps (which can be
associated with projected, thermal pressure maps), whereas the maps in the
bottom row are pressure maps derived based on X-ray surface brightness and
spectroscopic temperature maps, see equation \ref{eqn:p1} and
\ref{eqn:p2}. 
Both kinds of maps show an increase of structure (lumpiness) for the 
simulation which uses the
reduced viscosity scheme (right panels) when compared with the
original SPH viscosity scheme (left panels). The maps based on X-ray
observables show a larger degree of lumpiness due to the gas around
substructures, especially in the vicinity of infalling subgroups.}
\label{fig:pmaps}
\end{figure*}

\section{Cluster Properties} \label{sec:cluster}

Different levels of small-scale random gas motions within the ICM have only
mild effects on global properties of clusters like mass or temperature, as
evidenced by the measurements in Table~\ref{tab:char}.  However, additional
kinetic energy in turbulent motions changes the central density and entropy
structure, which in turn has a sizable effect on the X-ray luminosity.  We
investigate the resulting changes in cluster scaling relations and radial
profiles in more detail within this section.

\subsection{Maps}

The presence of a significant turbulent energy component in the
intra-cluster medium manifests itself in a modification of the balance
between the gravitational potential and the gas pressure.  There are
in fact observational reports that claim to have detected such
fluctuations in pressure maps derived from X-ray observations
\citep{2004A&A...426..387S}.  We here calculate artificial pressure
$P_{\rm art}$ maps for our simulations, based on surface brightness
maps ($L_x$) and spectroscopic-like \citep{2004MNRAS.354...10M}
temperature ($T_{sl}$) maps. This allows artificial pressure maps to
be estimated as
\begin{equation}
   P_{\rm art}=n\,T_{sl},
\label{eqn:p1}
\end{equation}
where
we defined
\begin{equation}
   n=\left(L_x/\sqrt{T_{sl}}\right)^{\frac{1}{2}}.
\label{eqn:p2}
\end{equation}

Figures~\ref{fig:lxtxmaps} and \ref{fig:pmaps} show a comparison
of a number of cluster maps produced using an unsharp-mask technique
of the form
\begin{equation}
\mathrm{Image}_\mathrm{unsharp mask} = \mathrm{Image}-\mathrm{Smoothed}(\mathrm{Image,\sigma}),
\end{equation}
where a Gaussian smoothing with FWHM of $\sigma=200$ kpc was applied. We
analyse maps of the X-ray surface brightness, spectroscopic-like temperature,
`true' pressure maps (e.g.~based on Compton $y$) and artificial pressure maps
constructed as described above. All maps show the central 2 Mpc of the cluster
run {\it g1}, simulated with the low--viscosity scheme (right panels)
compared with the original SPH scheme (left panels).

Disregarding the large contribution by substructure in all the X-ray
related maps (therefore also in the artificial pressure map), all
types of maps show clear signs of turbulence. It is noticeable in both
runs, but it has a much larger extent and intensity in the the
low--viscosity run. Note in particular the turbulent motions (appearing
as lumpiness in the unsharp-mask images) in the wake of infalling
substructures, and the earlier break-up of fluid interfaces when the
new, reduced viscosity scheme is used.

Pressure maps (and therefore SZ maps) are arguably the most promising
type of map when searching for observational imprints of
turbulence. Apart from reflecting the large scale shape of the cluster
they are known to be relatively featureless, because most of the
substructures in clusters are in pressure equilibrium with their local
surroundings, making them in principle invisible in pressure maps. On
the other hand, the contribution of the turbulent motion to the local
pressure balance can be expected to leave visible fluctuations in the
thermal pressure map. This can indeed be seen nicely in the pressure
(e.g. SZ) maps in Figure~\ref{fig:pmaps}. Note that the amplitudes of
the turbulent fluctuations in the case of the low--viscosity run are
larger and also spatially more extended in the core of the
cluster. Artificial pressure maps constructed from the X-ray
observables still show such fluctuations, but they are partially
swamped by the signatures of the infalling substructures, making it
difficult to quantify the amount of turbulence present in clusters
using such X-ray based artificial pressure maps.

The small displacements seen in the substructure positions between the
two runs are due to small differences in their orbital phases.
Besides the general problem to precisely synchronise cluster
simulations with different dynamical processes involved, it is well
known \citep[e.g.][]{2004MNRAS.350.1397T,2005astro.ph..4206P} that the
interaction of the gas with its environment can significantly change
the orbits of infalling substructure.  The different efficiencies in
stripping the gas from the infalling substructure in the simulations
with different viscosity prescription can therefore lead to small
differences in the timing and orbits between the two simulations.

\subsection{Scaling Relations}

\begin{figure}
\includegraphics[width=0.5\textwidth]{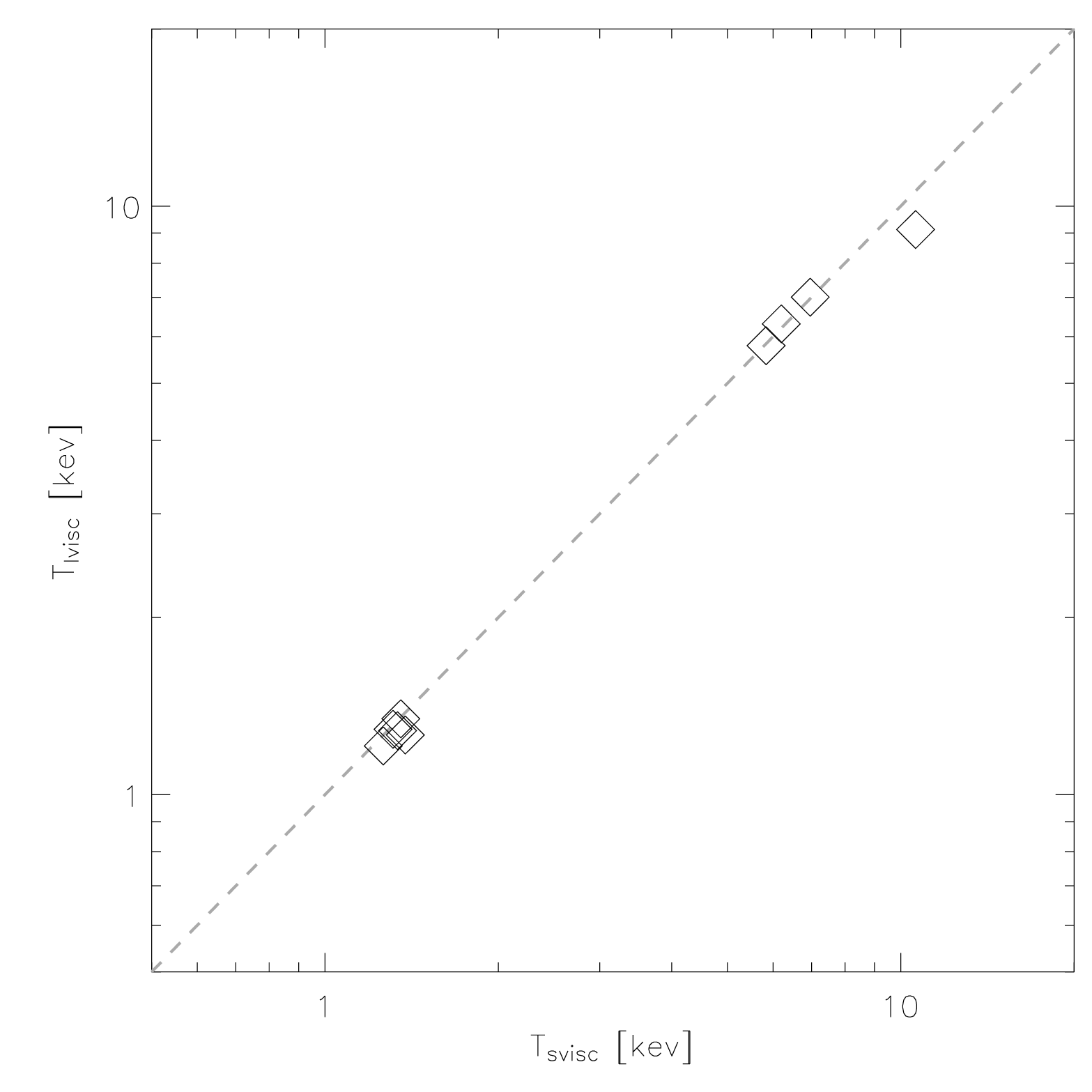}
\caption{Comparison of the virial temperature of
  the 9 clusters when different parameterisations of the viscosity are
  employed.  The solid line marks the one-to-one correspondence.}
\label{fig:t_t}
\end{figure}

\begin{figure}
\includegraphics[width=0.5\textwidth]{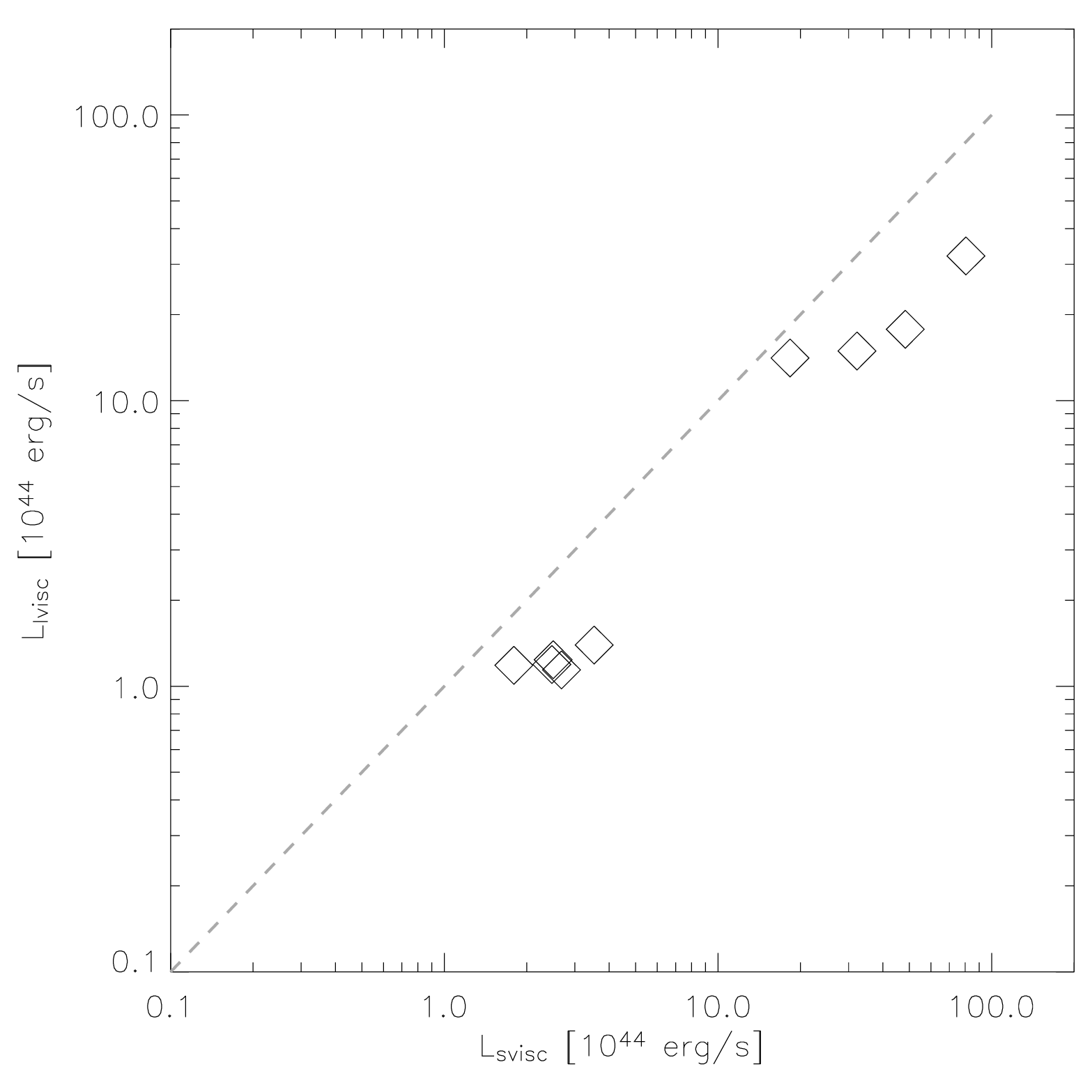}
\caption{Comparison of the bolometric luminosity of the 9 clusters
  when different parameterisations of the viscosity are employed. The
  solid line marks the one-to-one correspondence. It is evident that
  clusters with a larger degree of turbulence have a lower
  luminosity.}
\label{fig:lx_lx}
\end{figure}

In Figure~\ref{fig:t_t}, we compare the mass-weighted temperature of
our galaxy clusters for simulations with the original viscosity and
for runs with the low--viscosity scheme. There are no significant
changes.  Comparing the X-ray luminosity instead, we find that it
drops significantly by a factor of $\approx 2$ for clusters simulated
with the low--viscosity scheme, as shown in
Figure~\ref{fig:lx_lx}. This is quite interesting in the context of
the long-standing problem of trying to reproduce the observed cluster
scaling relations in simulations. In particular, since non radiative
cluster simulations tend to produce an excess of X-ray luminosity,
this effect would help.  However, one has to keep in mind that the
inclusion of additional physical processes like radiative cooling and
feedback from star formation can have an even larger impact on the
cluster luminosity, depending on cluster mass, so a definite
assessment of the scaling relation issue has to await simulations that
also include these effects.

\subsection{Radial profiles}

\begin{figure}
\includegraphics[width=0.5\textwidth]{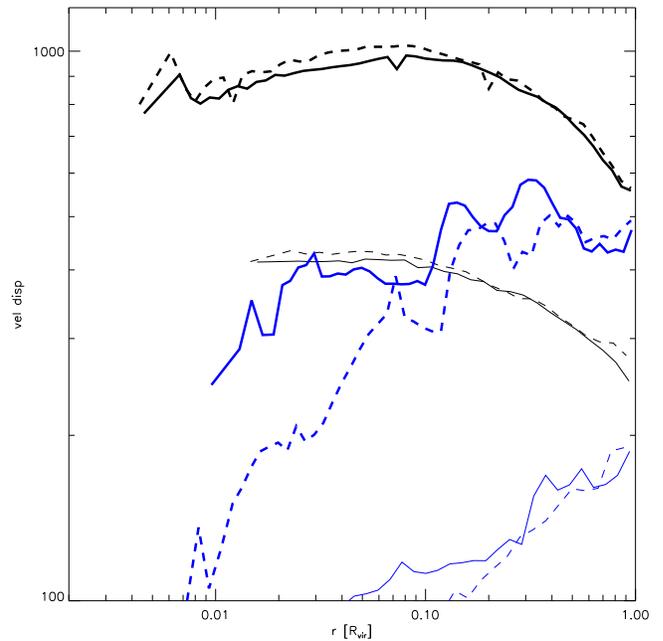}
\caption{Radial velocity dispersion profile for dark matter
  (black) and gas (blue) particles. The thick lines represent the average over
  the 4 massive clusters, whereas the thin lines give the average over the 5
  low mass systems. The dashed lines are drawn from the original viscosity
  simulations, the solid lines from the low--viscosity
  simulations.\label{fig:vdisp}}
\end{figure}

The presence of turbulence manifests itself in an increase of the velocity
dispersion of the cluster gas, especially towards the centre, while the dark
matter velocity dispersion should be unaffected.  In Figure~\ref{fig:vdisp},
we compare the velocity dispersion of gas and dark matter, averaged over the
low- and high-mass clusters in our set. As expected, the velocity dispersion
of the dark matter does not change in the low--viscosity simulations, where a
larger degree of turbulence is present in the ICM. On the other hand, the
central velocity dispersion of the gas increases, reaching of order of 
$400\,{\rm km\,s^{-1}}$ for our massive clusters. As the gas is in pressure
equilibrium with the unchanged gravitational potential, the hydrodynamic gas
pressure will be correspondingly lower in the centre due to the presence of
these random gas motions.

\begin{figure}
\includegraphics[width=0.5\textwidth]{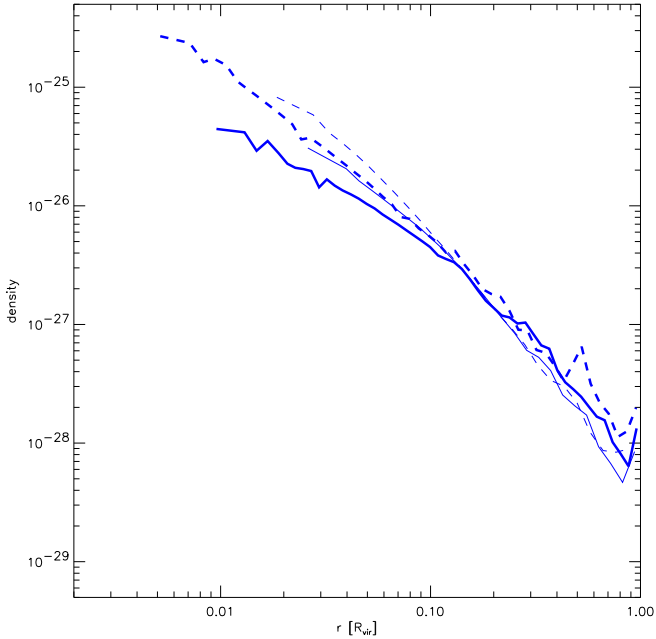}
\caption{Radial gas density profile. The thick lines
  represent the average over the 4 massive clusters, whereas the thin lines
  give the average over the 5 low mass systems. The dashed lines are drawn
  from the original viscosity simulations, the solid lines from the 
  low--viscosity simulations.\label{fig:dens_prof}}
\end{figure}

\begin{figure}
\includegraphics[width=0.5\textwidth]{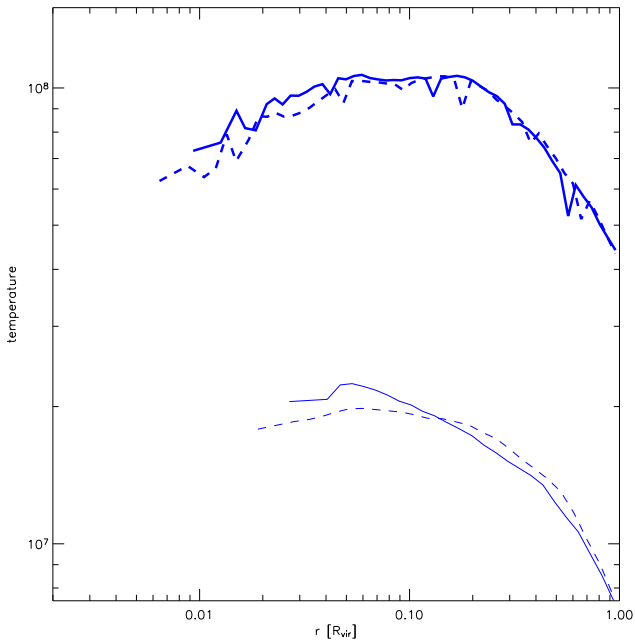}
\caption{Mass-weighted gas temperature profile. The thick lines
  represent the average over the 4 massive clusters, whereas the thin lines
  give the average over the 5 low mass systems. The dashed lines are drawn
  from the original viscosity simulations, the solid lines from the 
  low--viscosity simulations.\label{fig:temp_prof}}
\end{figure}

In Figure~\ref{fig:temp_prof}, we show the mean cluster temperature profiles,
which only shows a very mild trend of increasing temperature in
the central part of clusters when using the new, low--viscosity scheme.
However, the central gas density drops significantly in
the low--viscosity scheme, as shown in Figure~\ref{fig:dens_prof}. This
change in density is restricted to inner parts of the cluster, roughly to
within $~0.1R_{vir}$, which may be taken as an indication of the size of the
region where turbulent motions are important.

\begin{figure}
\includegraphics[width=0.5\textwidth]{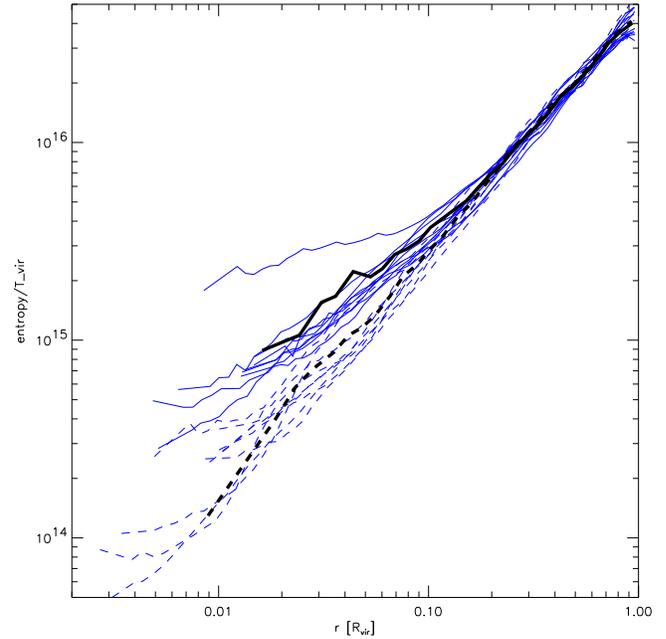}
\caption{Radial entropy profiles of the ICM gas.
  Thin lines are individual profiles for the 9 clusters, thick lines are
  averages.  The dashed lines are drawn from the original viscosity
  simulations, the solid lines from the low--viscosity
  simulations.\label{fig:entr_prof}}
\end{figure}

Quite interestingly, the presence of turbulence also changes the entropy
profiles. In Figure~\ref{fig:entr_prof}, we show the radial entropy profiles
of our clusters, which in the case of the low--viscosity scheme exhibit an
elevated level of entropy in the core, with a flattening similar to that
inferred from X-ray observations. It is remarkable that this central increase
of entropy occurs despite the fact that the source of entropy generation, the
artificial viscosity, is in principle less efficient in the low--viscosity
scheme. There are two main possibilities that could explain this result.
Either the low--viscosity scheme allows shocks to penetrate deeper into the
core of the cluster and its progenitors such that more efficient entropy
production in shocks occurs there, or alternatively, the reduced
numerical viscosity changes the mixing processes of infalling material,
allowing higher entropy material that falls in late to end up near the cluster
centre.

\begin{figure*}
\includegraphics[width=1.0\textwidth]{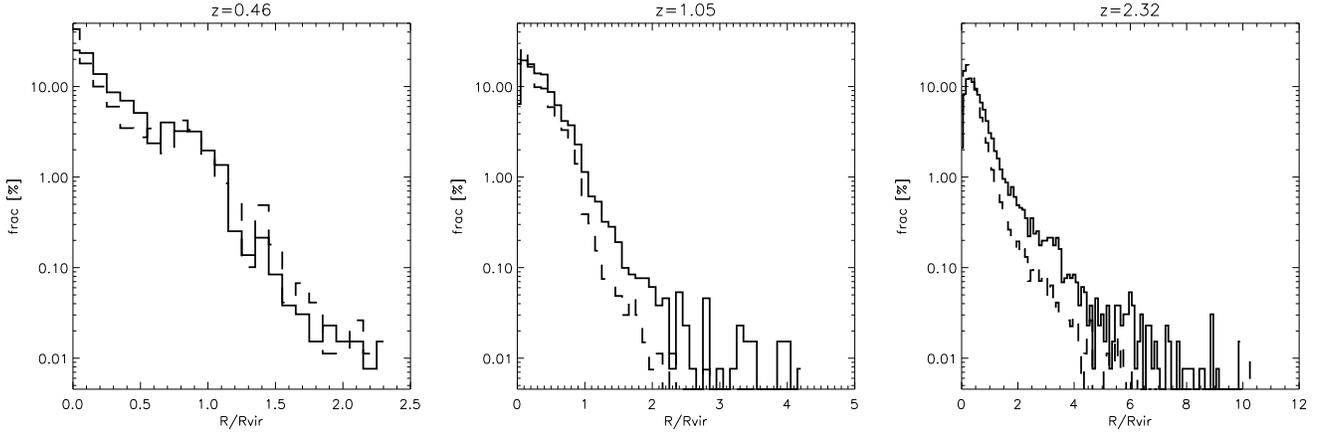}
\caption{ Distribution of the distance of particles to their nearest
  halo at high redshift.  The particles selected here end up within
  5\% of $R_{\rm vir}$ at $z=0$.  The dashed lines are for the original
  viscosity scheme, while the solid lines mark the result for the
  low--viscosity simulations.}
\label{fig:part_hist}
\end{figure*}

In order to investigate a possible change of the accretion behaviour, we traced
back to high redshift all particles that end up at $z=0$ within 5\% of
$R_{\rm vir}$ of the cluster centre.  We find that most of the central material is
located in the centres of progenitor halos at higher redshift, which is a well
known result.  However, in the simulations with the time dependend, low--viscosity scheme,
there is a clear increase of the number of particles which are not associated
with the core of a halo at higher redshift.  We illustrate this with the
histograms shown in Figure~\ref{fig:part_hist}, which gives the distribution of
the distance to the nearest halo in units of $R_{\rm vir}$ of the halo.  All
particles at distances larger than 1 are not associated with any halo at
corresponding epoch. Compared to the low entropy material that is already
bound in a dense core at this epoch, this diffuse gas is brought to much higher
entropy by shocks. When it is later accreted onto the cluster and mixed into
the core, it can then raise the entropy level observed there.  We note that
Eulerian hydrodynamics simulations also show a flattening of the entropy
profile. While the exact degree to which numerical and physical (turbulent)
mixing contribute to producing this result is still a matter of debate, it is
intriguing that a larger level of turbulence in the SPH calculations
substantially alleviates the discrepancies in the results otherwise obtained
with the two techniques
\citep{1999ApJ...525..554F,2003astro.ph.12651O}.

\section{Metal Lines} \label{sec:lines}

Turbulent gas motions can lead to substantial Doppler broadening of
X-ray metal lines, in excess of their intrinsic line widths. Given the
exquisite spectral resolution of upcoming observational X-ray mission,
this could be used to directly measure the degree of ICM turbulence
\citep{2003AstL...29..783S} by measuring, e.g., the shape of the
$6.702\,{\rm keV}$ iron emission line.

One potential difficulty in this is that multiple large-scale bulk
motions of substructure moving inside the galaxy cluster along the
line of sight might dominate the signal. To get a handle on this,
we estimate the line-of-sight emission of the $6.702\,{\rm keV}$
iron line within columns through the simulated clusters, where the
column size was chosen to be $300\,h^{-1}{\rm kpc}$ on a side,
which at the distance of the Coma cluster corresponds roughly to
one arcmin, the formal resolution of {\small ASTRO-E2}.  For
simplicity, we assign every gas particle a constant iron abundance
and an emissivity proportional to $n_e^2\times f(T_e) \times
\Delta V$, where $n_e$ is the electron density, and $\Delta
V\propto \rho^{-1}$ is the volume represented by the particle. As
a further approximation we set the electron abundance equal to
unity. We also neglect thermal broadening and other close lines
(like the $6.685\,{\rm keV}$ iron line), given that the
$6.702\,{\rm keV}$ iron line is clearly the strongest.

\begin{figure*}
\includegraphics[width=1.0\textwidth]{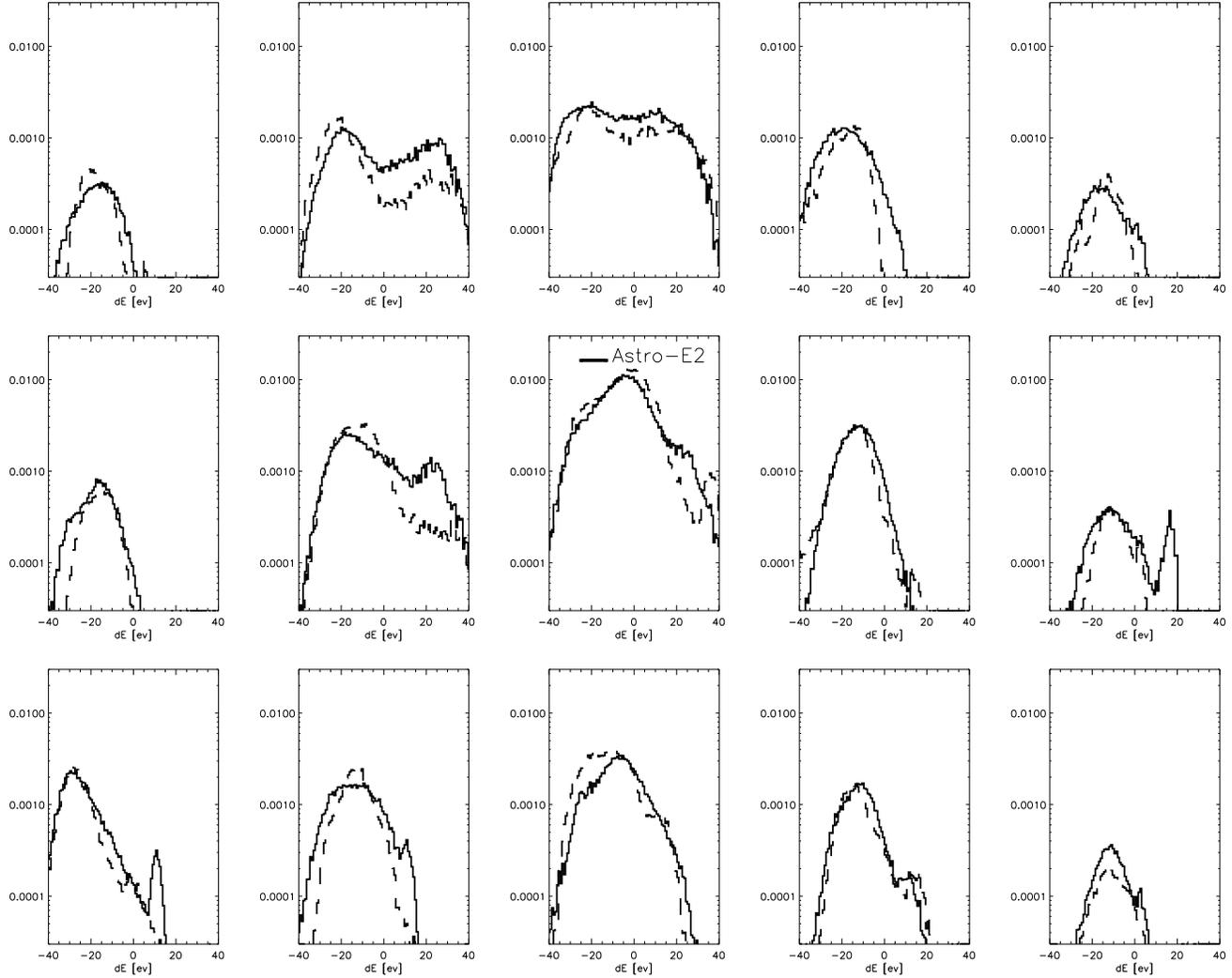}
\caption{Distribution of the Doppler-shifted
  emission of the iron $6.702\,{\rm keV}$ line for 15 lines of sight through
  the cluster {\it g72}. Every panel corresponds to a column of side length
  $300\,h^{-1}{\rm kpc}$ through the virial region of the cluster. This
  roughly corresponds to one arcmin resolution (comparable to the
  ASTRO-E2
  specifications) at the distance of the Coma cluster. The columns from left
  to right correspond to $-500$, $-250$, $0$, $250$, and $500\,h^{-1}{\rm
    kpc}$ impact parameter in the $x$-direction, and the rows correspond to
  $-250$, $0$, and $250,h^{-1}{\rm kpc} $ impact parameter in the
  $y$-direction. The dashed lines give results for the original viscosity run,
  while the solid line is for the low--viscosity run.  The thick bar in the
  center panel marks the expected energy resolution of 12 eV of the XRS
  instrument on-board ASTRO-E2.}
\label{fig:feline}
\end{figure*}

In Figure~\ref{fig:feline}, we show the resulting distributions for
several lines of sight, here distributed on a grid with $-500$,
$-250$, $0$, $250$ and $500\,h^{-1}{\rm kpc}$ impact parameter in
$x$-direction, and $-250$, $0$ and $250,h^{-1}{\rm kpc}$ impact
parameter in the $y$-direction, respectively. The different lines in
each panel correspond to simulations with the signal-velocity
based viscosity (dashed line) and with the time-depended low--viscosity
scheme (solid lines). Both results have been normalised to the total
cluster luminosity, such that the integral under the curves
corresponds to the fraction of the total luminosity.

We note that this measurement is very sensitive to small timing
differences between different simulations, and therefore a comparison
of the same cluster run with different viscosity should be carried out
in a statistical fashion, even if some features look very similar in
both runs. In general we confirm previous findings
\citep[e.g.][]{2003AstL...29..791I} that large bulk motions can lead
to spectral features which are several times wider than expected based
on thermal broadening alone. Additional complexity is added by beam
smearing effects, thermal broadening, and by the local turbulence in
the ICM gas, such that an overall very complex line shape results. In
our simulations with the low--viscosity scheme, where we have found
an increased level of fluid turbulence, the final line shapes are
indeed more washed out. However, the complexity of the final line
shapes suggests that it will be very difficult to accurately measure
the level of fluid turbulence with high resolution spectra of X-ray
emission lines, primarily because of the confusing influence of
large-scale bulk motions within galaxy clusters.

\section{Application to Radio Halos}
\label{sec:radio}
  One promising possibility to explain the extended radio emission on
  Mpc-scales observed in a growing number of galaxy clusters is to attribute
  it to electron acceleration by cluster turbulence
  \citep[e.g.,][]{1987A&A...182...21S,2001MNRAS.320..365B}.  Having high
  resolution cluster simulations at hand, which thanks to the new viscosity
  scheme are able to develop significant turbulence within the ICM, it is of
  interest to explore this possibility here.  Obviously, due to the
  uncertainties in the complex physical processes of dissipation of the
  turbulent energy -- which up to this point can not be explicitly modelled in
  the simulations -- our analysis is limited to a check whether or not
  turbulent reacceleration can plausibly reproduce some of the main properties
  of radio halos.  In this scenario, the efficiency of electron acceleration
depends on the energy density of magnetohydrodynamic waves (Alfv\'en waves,
Fast Mode waves, \ldots), on their energy spectrum, and on the physical
conditions in the ICM (i.e., density and temperature of the thermal plasma,
strength of the magnetic field in the ICM, number density and spectrum of
cosmic rays in the ICM). A number of approaches for studying the acceleration
of relativistic electrons in the ICM have been successfully developed by
focusing on the case of Alfv\'en waves
\citep{2002ApJ...577..658O,2004MNRAS.350.1174B} and, more recently, on Fast
Mode-waves \citep{2005MNRAS.357.1313C}.

It should be stressed, however, that analytical and/or
semi--analytical computations are limited to very simple assumptions
for the generation of turbulence in the ICM. Full numerical
simulations represent an ideal complementary tool for a more general
analysis, where the injection of turbulence into the cluster volume by
hierarchical merging processes can be studied realistically. Low
numerical viscosity and high spatial resolution are however a
prerequisite for reliable estimates of turbulence. As we have seen
earlier, previous SPH simulations based on original viscosity
parameterisations have suppressed random gas motions quite strongly,
but the low--viscosity scheme explored here does substantially better
in this respect. 

In this Section, we carry out a first exploratory analysis of the
efficiency of electron acceleration derived in the low--viscosity scheme,
and we compare it to results obtained with a original SPH
formulation. For definiteness, we assume that a fraction $\eta_t$ of
the estimated energy content of the turbulent velocity fields in
the cluster volume, measured by the local velocity dispersion
(equation \ref{sigmai}, section \ref{sec:turbulence}), 
is in the form of Fast Mode waves.  We focus on
these modes since relativistic electrons are mainly accelerated by
coupling with large scale modes (e.g., $k^{-1} \geq$ kpc, $k$
being the wave number) whose energy density, under the above
assumption, can hopefully be
constrained with the numerical simulations in a reliable fashion.
In addition, the damping and time evolution of Fast Modes
basically depend only on the properties of the thermal plasma and
are essentially not sensitive to the presence of cosmic ray
protons in the ICM \citep{2005MNRAS.357.1313C}.

Relativistic particles couple with Fast Modes via magnetic
Landau damping. The necessary condition for Landau damping
\citep{1968Ap&SS...2..171M,1979ApJ...230..373E} is $\omega -
k_{\Vert} v_{\Vert}=0$, where $\omega$ is the frequency of the
wave, $k_{\Vert}$ is the wavenumber projected along the magnetic
field, and $v_{\Vert}=v\mu$ is the projected electron velocity.
Note that in this case - in contrary to the Alfv\'enic case -
particles may also interact with large scale modes. 
In the collisionless regime, it can be shown that the resulting acceleration
rate in an isotropic plasma (modes' propagation and particle momenta) is given
by \citep[e.g.,][]{2005MNRAS.357.1313C}
\begin{equation}
{{ {\rm d} p }\over{{\rm d} t}} \sim 180 \,{{ v_{\rm M}^2
}\over{c}} {{p}\over{B^2}} \int
{\rm k} {W}^{B}_{k} {\rm d}k \label{dppms},
\label{dpdt}
\end{equation}
where $v_{\rm M}$ is the magneto--sonic velocity, and ${W}^{B}_{k}$ is
the energy spectrum of the magnetic field fluctuations
\citep[e.g.,][]{1973ApJ...184..251B,2005MNRAS.357.1313C}.  We estimate the
rate of injection of Fast Modes, $I^{FM}_k$, assuming that a fraction,
$\eta_t$, of fluid turbulence is associated with these modes. We parameterise
the injection rate assuming that turbulence is injected (and also dissipated)
in galaxy clusters within a time of the order of a cluster--crossing time,
$\tau_{\rm cross}$ \citep[see][for a more detailed
discussion]{2005MNRAS.357.1313C,2005astro.ph..5144S}. One then has:

\begin{equation}
\int I^{FM}_k \,{\rm d}k \sim \eta_t {{ E_t}\over{\tau_{\rm cross}}}
 \sim 
{1\over 2}
\eta_t \rho_{\rm gas} \sigma_v^2 \tau_{\rm inject}^{-1}\\
\label{inj}
\end{equation}

\noindent

Here, $\tau_{\rm inject}$ is the time over which a merging substructure is
able to inject turbulence in a given volume element in the main cluster.  This
can be estimated as the size of the subhalo divided by its infalling velocity.
As the size of a halo is only a weak function of its mass, we approximate
$\tau_{\rm inject}$ with a generic value of $\tau_{\rm inject}=0.5\,{\rm
  Gyr}$. This is only a very crude estimate and more generally one should
think of an effective efficiency parameter $\eta_t^{\rm eff}=\eta_t/\tau_{\rm
  inject}$ which we set to $0.1/(0.5\,{\rm Gyr})$ as argued before.  Note also
that for estimating $\sigma_v^2$ we used a $64^3$ TSC-grid, which is a
conservative estimate, as shown in Figure~\ref{fig:disp_resol}, and therefore
equation~(\ref{inj}) should still reflect a lower limit.

Following Cassano \& Brunetti (2005), the spectrum of the magnetic fluctuations
associated with Fast Modes is computed under stationary conditions taking into
account the damping rate of these modes with thermal electrons,
$\Gamma_k=\Gamma_o k$. One then has
\begin{equation}
W_k^B \sim {{B_o^2}\over{8 \pi}} {1\over{P_{\rm gas}}} 
{{I^{FM}_k}\over{\Gamma_o k}}.
\label{wbk}
\end{equation}

\noindent
Thus the integral in Eqn.~(\ref{dpdt}) at each position of the grid can be
readily estimated as
\begin{eqnarray}
\int k W_k^B\,{\rm d}k & \sim &
{{B_o^2}\over{8 \pi}} {1\over{ \Gamma_o P_{\rm gas}}}
\int I^{FM}_k\, {\rm d}k \\ 
& \sim & 
\eta_t^{\rm eff} {{ B^2({\bf x}) }\over{16 \pi}}
{{\rho_{\rm gas}({\bf x}) \sigma_{ii}^2({\bf x}) }\over {P_{\rm
gas}({\bf x})}}
{{ 1 }\over{\Gamma_o}}
\label{integral}
\end{eqnarray}
where $\Gamma_o$ depends on the
temperature of the ICM (Cassano \& Brunetti 2005)\footnote{Note that
under these assumptions the efficiency of the particle acceleration
does not depend on the spectrum of the waves}.

In this Section we are primarily interested in determining the maximum
energy of accelerated electrons, given the adopted energy density for
Fast Modes.  Under typical conditions in the ICM, the maximum energy
of electrons is reached at energies where radiative losses balance the
effect of the acceleration. The radiative synchrotron and inverse
Compton losses with CMB photons are given by
\citep[e.g.,][]{1999ApJ...520..529S}
\begin{eqnarray}
\lefteqn{ \left( {{{\rm d} p}\over{{\rm d} t}}\right)_{\rm rad}=\,
-4.8\times10^{-4} p^2 \left[\left({{ B_{\mu G}}\over{
3.2}}\right)^2{{\sin^2\theta}\over{2/3}} +(1+z)^4 \right] }{}
\nonumber\\
& & {} \,\,\,\,\,\,\,\,\,\,\,\,\,\,\,\,\,\ = -\frac{\beta\,
p^{2}}{m_e\,c} \label{syn+ic} \label{losses},
\end{eqnarray}
where $B_{\mu G}$ is the magnetic field strength in $\mu G$, and $\theta$ is
the pitch angle of the emitting electrons. If an efficient isotropisation of
electron momenta can be assumed, it is possible to average over all possible
pitch angles, so that $\left<\sin^2\theta\right> = 2/3$.

\begin{figure}
\includegraphics[width=0.5\textwidth]{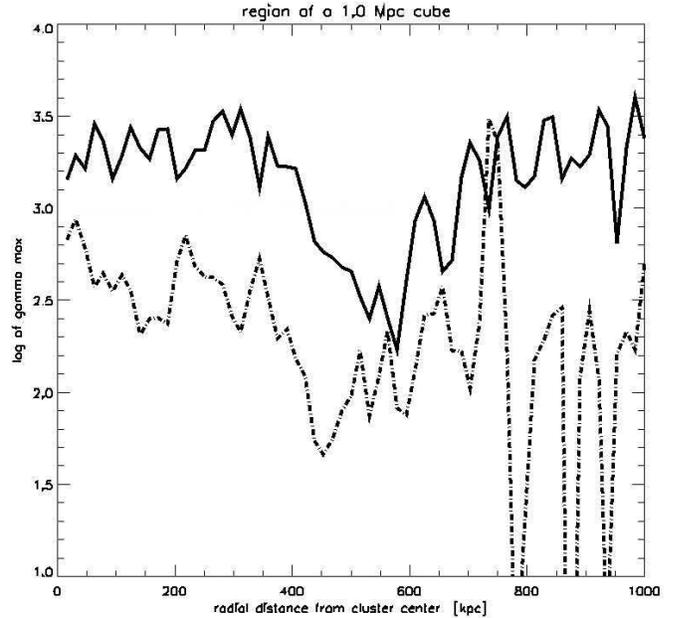}
\caption{One-dimensional profile of the maximum energy of
  the electrons accelerated via the turbulent-magneto-sonic model, along the
  same vertical lines drawn in Figure~\ref{fig:vel_mean}.  Dashed lines are
  for the original viscosity run, while solid lines are for the 
  low--viscosity
   scheme. Here, a conservative $64^3$ grid is used in the TSC
    smoothing.}
\label{fig:radio1d}
\end{figure}

In Figure~\ref{fig:radio1d}, we plot the maximum energy of the fast electrons
obtained from Eqs.~(\ref{wbk}) and (\ref{losses}) along one line-of-sight
through the cluster atmosphere. The two different lines are for the same
cluster, simulated with our two main schemes for the artificial viscosity. The
two vertical lines in Figure~\ref{fig:vel_mean} are indicating the position of
these cuts.  When the new low--viscosity scheme is used, enough turbulence is
resolved to maintain high energy electrons (and thus synchrotron radio
emission) almost everywhere out to a distance of 1 Mpc from the cluster
centre, whereas in the original formulation of SPH, turbulence is much more
strongly suppressed, so that the maximum energy of the accelerated electrons
remains a factor of about three below that in the low--viscosity case.

The results reported in Figure~\ref{fig:radio1d} are obtained assuming a
reference value of $\eta_t^{\rm eff}=\eta_t/\tau_{\rm inject}=0.1/(0.5\,{\rm
  Gyr})$.  The averaged volume weighted magnetic field strength in the
considered cluster region is fixed at $0.5\mu G$ and a simple scaling from
magnetic flux--freezing, $B\propto\rho^{(2/3)}$, is adopted in the
calculations, resulting in a central magnetic field strength of $B_0=5.0\mu
G$. It is worth noting that the maximum energy of the accelerated electrons,
$\gamma_{\rm max}$, scales with the energy density of the turbulence (and with
the fraction of the energy of this turbulence channelled into Fast Modes
$\eta_t$, Eq.~\ref{wbk}). However the synchrotron frequency emitted by these
electrons scales with the square of the turbulent energy ($\gamma_{\rm
  max}^2$).  Interestingly, with the parameter values adopted in
Figure~\ref{fig:radio1d}, a maximum synchrotron frequency of order 10$^2(
{{\eta_t}/{0.1}} )^2$ MHz is obtained in a Mpc--sized cluster region, which
implies diffuse synchrotron radiation up to GHz frequencies if a slightly
larger value of $\eta_t$ is adopted\footnote{Note that
  \citep[e.g.,][]{2005MNRAS.357.1313C} required $\eta_t \sim 0.2-0.3$ in order
  to reproduce the observed statistics of radio halos.}. On the other hand, we
note that essentially no significant radio emission would be predicted if we
had used the simulations with the original SPH viscosity scheme.

In real galaxy clusters, the level of turbulence which can form will
also depend on the amount of physical viscosity present in the ICM gas
(i.e.~on its Reynolds number), which in turn depends on the magnetic
field topology and gas temperature.  It will presumably still take a
while before the simulations achieve sufficient resolution that the
numerical viscosity is lower than this physical viscosity.  In
addition, the details of the conversion process of large--scale
velocity fields into MHD modes is still poorly understood and well
beyond the capabilities of presently available cosmological
simulations.  However, our results here show that a suitable
modification of the artificial viscosity parameterisation within SPH
can be of significant help in this respect, and it allows a first
investigation of the role of turbulence for feeding non--thermal
phenomena in galaxy clusters.

\section{Conclusions} \label{sec:conclusions}

We implemented a new parameterisation of the artificial viscosity of
SPH in the parallel cosmological simulation code {\small
GADGET-2}. Following a suggestion by \cite{1997JCoPh..136....41S},
this method amounts to an individual, time-dependent strength of the
viscosity for each particle which increases in the vicinity of shocks
and decays after passing through a shock.  As a result, SPH should
show much smaller numerical viscosity in regions away from strong
shocks than original formulations.  We applied this low--viscosity
formulation of SPH to a number of test problems and to cosmological
simulations of galaxy cluster formations, and compared the results to
those obtained with the original SPH formulation.  
Our main results can be summarised as follows:

\begin{itemize}

\item The low--viscosity variant of SPH is able to capture strong shocks just
  as well as the original formulation, and in some cases we even obtained
  improved results due to a reduced broadening of shock fronts. In spherical
  accretion shocks, we also obtained slightly better results due to a
  reduction of pre-shock entropy generation.

\item Using the low--viscosity scheme, simulated galaxy clusters
  developed significant levels of turbulent gas motions, driven by
  merger events and infall of substructure.  We find that the kinetic
  energy associated with turbulent gas motion within the inner $\sim
  1\,{\rm Mpc}$ of a $10^{15}\,h^{-1}{\rm M}_\odot$ galaxy cluster can
  be up to 30\% of the thermal energy content.  This value can be
  still larger and reach up to 50\% in the very central part of
  massive clusters. In clusters with smaller masses ($\sim 10^{14}
  \,h^{-1}{\rm M}_\odot$) we find a smaller turbulent energy content,
  reaching only 5\% within the central Mpc. Within a comparable
  fraction of the virial radius, the corresponding fraction is however
  still of order 10\%.  These values are much larger than what is
  found when the original SPH viscosity is employed, which strongly
  suppresses turbulent gas motions.  

\item The presence of such an amount of turbulence has an imprint on global
  properties of galaxy clusters, most notably reducing the bolometric
  X-ray luminosity in non radiative simulations by a factor of $\approx 2$.
  However, the global, mass-weighted temperature does not change.

\item The temperature profiles of galaxy clusters are only mildly
  changed by the presence of turbulence, but we observe a strong decrease of
  density within the central region of galaxy clusters, where the turbulence
  is providing a significant contribution to the total pressure. Also the
  radial entropy profiles show a significant flattening towards the cluster
  centre. This makes them more similar to the observed profiles based on X-ray
  observations. Note however that radiative cooling -- which was not included
  in our simulations -- can also modify the profiles substantially.  We find
  that the higher entropy in the centre found in the low viscosity simulations
  is largely a result of the more efficient transport and mixing of
  low-entropy in infalling material into the core of the cluster. We note that
  the elevated entropy levels found in our low--viscosity runs are more similar
  to the results found with Eulerian hydrodynamic codes than the 
  original SPH ones.

\item Turbulence in galaxy clusters broadens the shape of metal lines
  observable with high-resolution X-ray spectrographs like XRT on board of
  {\small ASTRO-E2}. Depending on the strength of the turbulence and the
  dynamical state of the cluster, prominent features due to large-scale bulk
  motions may however get washed out and blended into a very complex line
  structure.  In general it will therefore be difficult to isolate the
  signature of the turbulent broadening and to differentiate it unambiguously
  from the more prominent features of large scale bulk motions.

\item Applying a model for accelerating relativistic electrons by ICM
  turbulence we find that galaxy clusters simulated with reduced
  viscosity scheme may develop sufficient turbulence to account for
  the radio emission that is observed in many galaxy clusters,
  provided that a non--negligible fraction of the turbulent energy in
  the real ICM is associated with Fast Modes.
\end{itemize}

In summary, our results suggest that ICM turbulence might be an
important ingredient in the physics of galaxy clusters. If present at
the levels inferred from our low--viscosity simulations, it has a
significant effect on the radial structure and on the scaling
relations of galaxy clusters. We also note that the inferred reduction
of the X-ray luminosity has a direct influence on the strength of
radiative cooling flows. The more efficient mixing processes would
also help to understand the nearly homogeneous metal content observed
for large parts of the cluster interior. Finally, cluster turbulence
may also play an important role for the dynamics of non-thermal
processes in galaxy clusters.

Although we observe a rather high level of turbulence in the very centre of
our simulated galaxy clusters when we use the low--viscosity scheme, it is
likely that we are still missing turbulence due to the remaining numerical
viscosity of our hydrodynamical scheme, and due to the resolution limitations,
particularly in low density regions, of our simulations.  This problem should
in principle become less and less severe as the resolution of the simulations
is increased in future calculations. However, given that there is a some
physical viscosity in real galaxy clusters which limits the Reynolds number of
the ICM, it cannot be the goal to model the ICM as a pure ideal gas. Instead,
future work should concentrate on accurately characterising this physical
viscosity of the ICM, which could then be directly incorporated into the
simulations by means of the Navier-Stokes equations. Our results suggest that
the low--viscosity formulation of SPH should be of significant help in
reducing the numerical viscosity of SPH simulation below the level of this
physical viscosity, and the present generation of simulations may already be
close to this regime.


\section*{acknowledgements}
Many thanks to Volker Springel for providing {\small GADGET-2} and
initial conditions for test problems. 
We acknowledge fruitful discussions with Stefano Borgani and want to
thank Volker Springel and Torsten Ensslin for carefully reading
and fruitful suggestions to improve the manuscript. The
simulations were carried out on the IBM-SP4 machine at the
``Centro Interuniversitario del Nord-Est per il Calcolo
Elettronico'' (CINECA, Bologna), with CPU time assigned under an
INAF-CINECA grant, on the IBM-SP3 at the Italian Centre of
Excellence ``Science and Applications of Advanced Computational
Paradigms'', Padova and on the IBM-SP4 machine at the
``Rechenzentrum der Max-Planck-Gesellschaft'' in Garching.
K.~D.~acknowledges support by a Marie Curie Fellowship of the
European Community program "Human Potential" under contract number
MCFI-2001-01227. G.~B.~acknowledges partial support from MIUR
through grant PRIN2004 and from INAF through grant D4/03/15.

\bibliographystyle{mnras}
\bibliography{master}

\end{document}